# Metamaterials in Superconducting and Cryogenic Quantum Technologies

Alex Krasnok[1,2]

[1]*Department of Electrical Engineering, Florida International University, 33174, Miami, USA*

[2]*Knight Foundation School of Computing and Information Sciences, Florida International University, 33174, Miami, USA*

To whom correspondence should be addressed: akrasnok@fiu.edu

## Abstract

The development of fault-tolerant quantum computers based on superconducting circuits faces critical challenges in qubit coherence, connectivity, and scalability. This review establishes metamaterials – artificial structures with on-demand electromagnetic properties – as a transformative solution. By engineering the photonic density of states, metamaterials can suppress decoherence via the Purcell effect and create multi-mode quantum buses for hardware-efficient control and long-range qubit coupling. We provide a comprehensive overview, from foundational principles and Hamiltonian engineering to the materials science of high-coherence devices. We survey state-of-the-art performance, highlighting record coherence times and coupling strengths achieved through metamaterial design. Furthermore, we explore advanced applications where engineered environments give rise to exotic excitations and topologically protected states, enabling novel error correction schemes and qubit architectures. Ultimately, we argue that metamaterials are evolving from passive components into the core architectural element of next-generation quantum technologies, paving a viable path toward scalable quantum computation.

## 1. Introduction

The pursuit of a large-scale, fault-tolerant quantum computer represents one of the most ambitious and transformative scientific endeavors of our time. Such a machine promises to solve certain classes of problems – in fields ranging from materials science and drug discovery to cryptography and complex optimization – that are intractable for even the most powerful classical supercomputers[1–6]. Among the various physical platforms being developed to realize this vision, superconducting quantum circuits have emerged as a leading contender, primarily due to their compatibility with well-established semiconductor fabrication techniques, which allows for the design of complex, integrated processors, and their demonstrated potential for scalability[7–10]. Despite remarkable progress over the past two decades, which has seen qubit coherence times increase by over five orders of magnitude and processor sizes grow to hundreds of qubits, the path to fault tolerance is fraught with fundamental challenges[11]. Superconducting qubits, while powerful, are exquisitely sensitive to their environment. Their fragile quantum states are susceptible to decoherence – the irreversible loss of quantum information – arising from a multitude of noise sources. This inherent fragility necessitates the use of quantum error correction (QEC), where logical information is redundantly encoded across many physical qubits to protect it from errors[12–15]. However, the resource overhead for conventional QEC schemes like the surface code is immense, requiring potentially thousands of physical qubits to create a single, robust logical qubit.

The central challenges hindering the development of large-scale superconducting quantum processors can be broadly categorized into three interconnected areas: coherence, connectivity, and control.

- **Coherence:** The lifetime of a qubit is fundamentally limited by energy relaxation and dephasing[16–22]. A dominant source of these errors is dielectric loss from microscopic defects, known as two-level systems (TLS), which are ubiquitous in the amorphous oxides and interfaces of the fabricated device. As device complexity increases, so does the number of interfaces, exacerbating the impact of these material-based loss channels.

- **Connectivity:** Many powerful quantum algorithms and efficient QEC codes require interactions between qubits that are not physically adjacent. Conventional processor architectures, which often feature only fixed, nearest-neighbor coupling, are therefore highly inefficient, requiring long sequences of error-prone SWAP gates to entangle distant qubits[23–26].

- **Control and Scalability:** Perhaps the most daunting practical challenge is the "wiring bottleneck"[27–35]. In current designs, each qubit requires multiple dedicated microwave control and readout lines that must be routed from room-temperature electronics into the cryogenic environment. This approach scales poorly, imposing a significant thermal load on the refrigerator and creating a physical barrier to integrating the millions of qubits required for fault tolerance.

Addressing these deeply rooted challenges requires more than just incremental improvements to existing designs; it demands a fundamental rethinking of the electromagnetic environment in which qubits operate. This is precisely the opportunity offered by the integration of *metamaterials* into superconducting quantum circuits[36–46]. Metamaterials are artificial, engineered structures whose electromagnetic properties arise not from their constituent materials but from their sub-wavelength geometric design[38,47–50]. This design paradigm provides an unprecedented level of control over the propagation of microwave photons, allowing for the creation of bespoke electromagnetic environments tailored to enhance quantum performance.

By constructing resonators from periodic arrays of lumped-element inductors and capacitors, it becomes possible to engineer the dispersion relation – the relationship between a photon's frequency and its momentum – in ways that directly address the challenges of coherence and scalability. For instance, creating a photonic bandgap can dramatically alter the density of states available for a qubit to decay into, suppressing spontaneous emission via the Purcell effect and thereby extending coherence times. Furthermore, by implementing unconventional dispersion, such as in left-handed metamaterials, one can create a dense spectrum of resonant modes within a compact physical footprint[23,43,51,51–53]. These multi-mode structures can serve as a quantum bus, mediating programmable, long-range interactions between distant qubits to overcome the connectivity problem, while also enabling frequency-multiplexed control schemes that alleviate the wiring bottleneck.

This review provides a comprehensive overview of the rapidly advancing field of superconducting metamaterial resonators for quantum technologies. We begin by establishing the foundational principles that make superconductors the ideal platform for quantum metamaterials. We then delve into the design methodologies for engineering the dispersion and impedance of these artificial structures. A detailed discussion of the materials science and fabrication challenges is presented, highlighting the critical link between microscopic defects and macroscopic quantum performance. We survey the state-of-the-art, presenting record values for key performance metrics that demonstrate the tangible benefits of the metamaterial approach. Finally, we explore the exciting frontiers where these engineered environments are enabling the realization of exotic physical phenomena – from topologically protected states to non-radiating anapole modes and artificial skyrmions – and are forming the backbone of advanced, hardware-efficient quantum error correction architectures. Through this exploration, a clear picture emerges: metamaterials are evolving from passive components into the core architectural element of next-generation superconducting quantum technologies.

## 2. Foundational Principles of Superconducting Metamaterials in Quantum Circuits

The unique properties of superconducting materials, when combined with the design principles of metamaterials, create a powerful platform for quantum technologies[36,37]. This synergy is not merely an incremental improvement over classical metamaterials but an enabling factor for realizing quantum-coherent, tunable, and compact devices that are otherwise unattainable. The field of metamaterials is built on the principle of engineering artificial materials with electromagnetic properties not found in nature, achieved by structuring them with periodic, sub-wavelength elements or "meta-atoms". When these meta-atoms are constructed from superconductors and operated at cryogenic temperatures, they inherit a suite of quantum mechanical properties that are exceptionally well-suited for the demands of quantum information processing. This section establishes the foundational physics that makes this combination so potent, answering the fundamental questions of "Why superconductors?" and "Why metamaterials for quantum circuits?".

### 2.1 The Quantum Advantage of Superconductivity: Low Loss and High Kinetic Inductance

At the heart of any quantum technology lies the challenge of preserving delicate quantum states against the relentless onslaught of environmental noise and energy dissipation. For quantum circuits operating in the microwave frequency domain, the choice of material is therefore paramount. Superconductivity, a macroscopic quantum state of matter, offers a near-perfect solution to the problem of dissipation[37,54–57]. The electrodynamic response of any material to an alternating electromagnetic field can be described by its complex conductivity, given by: $\sigma(\omega) = \sigma_1(\omega) - i\sigma_2(\omega)$. In this expression, the real part, $\sigma_1$, quantifies the dissipative losses in the material, primarily due to the scattering of charge carriers, which converts electromagnetic energy into heat. The imaginary part, $\sigma_2$, represents the reactive, inductive response arising from the inertia of the charge carriers. In a normal metal like copper or gold, a large population of free electrons leads to frequent scattering events, resulting in a large $\sigma_1$ and significant ohmic losses. This problem is severely exacerbated in metamaterial structures, where electromagnetic fields are often highly concentrated in sub-wavelength resonant elements, making conventional metals unsuitable for high-performance applications[58,59].

Superconductors fundamentally alter this picture. Below a critical temperature, $T_c$, electrons near the Fermi level bind together to form Cooper pairs. These pairs condense into a single, coherent macroscopic quantum state described by a complex order parameter[60]. A key consequence of this condensation is the opening of a superconducting energy gap, $2\Delta$, in the single-particle excitation spectrum. For quantum circuits operating at microwave frequencies (typically 3-15 GHz) and cryogenic temperatures ($T \ll T_c$), the energy of a microwave photon, $\hbar\omega$, is much smaller than the energy required to break a Cooper pair ($\hbar\omega \ll 2\Delta$). Consequently, the population of dissipative, unpaired electrons, known as quasiparticles, is exponentially suppressed. This leads to the defining electrodynamic property of a superconductor in the quantum regime: the dissipative component of its conductivity becomes vanishingly small compared to its inductive component, such that $\sigma_1 \ll \sigma_2$. The material behaves as an almost perfectly lossless inductor. This intrinsic low-loss nature is a critical advantage, as many of the exotic phenomena promised by metamaterials, such as negative refraction and evanescent wave amplification, are strongly suppressed by even small amounts of material loss[61,62].

Beyond being low-loss, the inductive response of a superconductor provides a powerful and unique design tool known as kinetic inductance. The total inductance of a superconducting wire has two components: the familiar geometric inductance, determined by the physical layout of the wire, and a kinetic inductance, $L_k$, which arises from the inertial mass of the Cooper pairs as they accelerate and decelerate in the microwave field[63–66]. The kinetic inductance per square of a thin superconducting film is given by the expression:

$$L_k = \frac{\hbar R_s}{\pi \Delta}, \quad (1)$$

where $R_s$ is the film's normal-state sheet resistance, $\hbar$ is the reduced Planck constant, and $\Delta$ is the superconducting energy gap[67–70]. This formula reveals a remarkable and *counter-intuitive design principle*: materials that are poor conductors in their normal state (i.e., have a high $R_s$), such as disordered or granular thin films of Niobium Nitride (NbN), Titanium Nitride (TiN), or aluminum, become exceptional "superinductors" when cooled into their superconducting state. This provides a powerful, non-geometric source of inductance that can be orders of magnitude larger than the geometric inductance.

The availability of high kinetic inductance is a cornerstone of superconducting metamaterial design. It allows for the fabrication of extremely compact resonant elements with dimensions deep in the sub-wavelength regime[37]. For instance, self-resonant superconducting spirals can be fabricated to operate at radio frequencies while being over 1000 times (!) smaller than the free-space wavelength of their fundamental mode, all while maintaining a high quality factor (Q). This extreme miniaturization, enabled by the combination of low loss and high kinetic inductance, is a fundamental advantage that distinguishes superconducting metamaterials from their normal-metal counterparts and makes them an ideal platform for building scalable quantum circuits.

## 2.2 The Josephson Junction as a Tunable, Nonlinear Meta-Atom

While low-loss linear elements are essential for storing quantum information, the ability to create, manipulate, and read out quantum states requires a crucial additional ingredient: nonlinearity[71]. In superconducting circuits, this is provided by the Josephson junction (JJ), a remarkable quantum device that serves as the cornerstone for virtually all superconducting qubits and tunable metamaterials. A Josephson junction consists of two superconducting electrodes separated by a nanometer-thin insulating barrier, such as aluminum oxide[72,73]. This barrier is thin enough to allow Cooper pairs to tunnel coherently from one superconductor to the other, a dissipation-free process known as the Josephson effect. The quantum dynamics of this tunneling process are described by two fundamental equations, the Josephson relations:

$$I(t) = I_c \sin(\delta(t)) \quad (2)$$

$$V(t) = \frac{\Phi_0}{2\pi} \frac{d\delta}{dt} \quad (3)$$

Here, $I(t)$ is the supercurrent flowing through the junction, $V(t)$ is the voltage across it, and $\delta(t)$ is the gauge-invariant quantum mechanical phase difference of the superconducting order parameter across the barrier. The parameter $I_c$ is the junction's critical current, which is the maximum supercurrent it can sustain without developing a voltage. The constant $\Phi_0 = h/2e \approx 2.067 \times 10^{-15}$ Wb is the magnetic flux quantum, a fundamental constant of nature that emerges from the charge of a Cooper pair ($2e$).

These equations reveal that the Josephson junction behaves as a perfect, non-dissipative inductor whose inductance is nonlinear; that is, its value depends on the current flowing through it. The Josephson inductance, $L_J$, can be derived from the relations above, $L_J(\delta) = \frac{V(t)}{dI(t)/dt} = \frac{\Phi_0/(2\pi) \cdot (d\delta/dt)}{I_c \cos(\delta) \cdot (d\delta/dt)} = \frac{\Phi_0}{2\pi I_c \cos(\delta)}$. The dependence of $L_J$ on $\cos(\delta)$ is the source of the nonlinearity. This property is absolutely essential for creating a qubit. A standard linear LC circuit is a harmonic oscillator, and its quantum mechanical energy levels are equally spaced. It is impossible to isolate a two-level system from such a ladder of states, because any microwave signal that drives the transition from the ground state to the first excited state would also be resonant with all higher-level transitions. The nonlinear inductance of the JJ creates an anharmonic potential energy landscape, described by $U_J = -E_J \cos(\delta)$, where $E_J = I_c \Phi_0/2\pi$ is the Josephson energy. When shunted by a capacitor with charging energy $E_C = (2e)^2/2C$, the resulting circuit is an anharmonic oscillator with unequally spaced energy levels. This anharmonicity allows one to use microwaves to selectively address the transition between the two lowest energy states, $|0\rangle$ and $|1\rangle$, thus realizing a qubit.

Furthermore, the properties of the Josephson junction can be made exquisitely tunable by embedding one or more junctions into a superconducting loop, thereby forming a Superconducting Quantum Interference Device (SQUID)[74,75]. The behavior of a SQUID is governed by the principle of magnetic flux quantization, which requires that the total magnetic flux enclosed by a superconducting loop must be an integer multiple of $\Phi_0$. This quantum constraint links the phase difference $\delta$ across the junction(s) to any externally applied magnetic flux, $\Phi_{ext}$, that threads the loop. Consequently, by applying a small, external DC magnetic field, one can precisely control the effective Josephson inductance of the SQUID.

This flux-based tuning mechanism is a defining advantage of SQUID-based metamaterials. It is orders of magnitude faster and more efficient than alternative tuning methods, such as varying the temperature or the applied current, which are often slow or introduce significant dissipation[37,40,76–82]. SQUIDs can therefore be viewed as nearly ideal, highly tunable meta-atoms. Arrays of SQUIDs can be used to create artificial one-dimensional crystals with tunable band structures, or to modulate the effective speed of light in a transmission line for applications in parametric amplification[37,82–84]. The combination of low loss from the superconductor, essential nonlinearity from the Josephson effect, and high-speed tunability from the SQUID geometry provides a complete and powerful toolkit for engineering the quantum electromagnetic environment.

## 2.3 Hamiltonian Formalism for Superconducting Circuits

A powerful and defining aspect of the superconducting circuit platform is the existence of a systematic and rigorous procedure to translate the classical description of a circuit, familiar from electrical engineering, into a full quantum mechanical Hamiltonian[85]. This process, known as *circuit quantization*, provides the crucial link between the physical layout of inductors and capacitors on a chip and the quantum phenomena – such as discrete energy levels, superposition, and entanglement – that emerge at cryogenic temperatures. It allows designers to engineer specific quantum behaviors by carefully choosing the circuit topology and component values, making it an indispensable tool for creating bespoke quantum metamaterials.

The quantization procedure begins with a classical description of the circuit's energy. Instead of using voltages and currents directly, it is more convenient to work with generalized coordinates and their corresponding velocities. For electrical circuits, the natural choice for a generalized coordinate is the *node flux*, $\Phi_n(t)$, which is the time integral of the voltage at a given node $n$, $\Phi_n(t) \equiv \int_{-\infty}^{t} V_n(\tau) d\tau$. The corresponding generalized velocity is the node voltage itself, $\dot{\Phi}_n(t) = V_n(t)$. Using these variables, one can construct the Lagrangian of the circuit, $\mathcal{L}$, which is the difference between the system's kinetic energy (energy stored in capacitors) and potential energy (energy stored in inductors and Josephson junctions).

Once the Lagrangian is established, the canonical momentum conjugate to each flux variable $\Phi_n$ is found. This conjugate momentum turns out to be the charge, $Q_n$, on the capacitor associated with that node: $Q_n = \frac{\partial \mathcal{L}}{\partial \dot{\Phi}_n}$. With the set of conjugate coordinate-momentum pairs $(\Phi_n, Q_n)$ identified, a Legendre transformation is performed to move from the Lagrangian to the Hamiltonian formalism: $\mathcal{H} = \sum_n Q_n \dot{\Phi}_n - \mathcal{L}$. The final step is the canonical quantization, where the classical variables are promoted to quantum operators, $\Phi_n \to \hat{\Phi}_n$ and $Q_n \to \hat{Q}_n$. These operators are constrained to obey the canonical commutation relation, analogous to the position and momentum operators in quantum mechanics: $[\hat{\Phi}_n, \hat{Q}_m] = i\hbar \delta_{nm}$. This procedure transforms the classical energy function of the circuit into a quantum Hamiltonian operator, $\hat{H}$, whose eigenvalues correspond to the quantized energy levels of the system and whose structure dictates the system's dynamics[85].

Let us apply this formalism to the simplest and most fundamental building block of a metamaterial: a linear LC resonator, which consists of an inductor $L$ in parallel with a capacitor $C$. The energy stored in the capacitor (kinetic energy) is $E_C = Q^2/2C = \frac{1}{2}C\dot{\Phi}^2$, and the energy stored in the inductor (potential energy) is $E_L = \Phi^2/2L$. The Lagrangian is therefore, $\mathcal{L} = E_C - E_L = \frac{1}{2}C\dot{\Phi}^2 - \frac{1}{2L}\Phi^2$. The canonical momentum is the charge $Q = \partial\mathcal{L}/\partial\dot{\Phi} = C\dot{\Phi}$. Performing the Legendre transformation gives the classical Hamiltonian, $\mathcal{H} = Q\dot{\Phi} - \mathcal{L} = \frac{Q^2}{C} - \left(\frac{Q^2}{2C} - \frac{\Phi^2}{2L}\right) = \frac{Q^2}{2C} + \frac{\Phi^2}{2L}$. Upon quantization, this becomes the Hamiltonian for a quantum harmonic oscillator[86,87]:

$$\hat{H}_{res} = \frac{\hat{Q}^2}{2C} + \frac{\hat{\Phi}^2}{2L} \quad (4)$$

This Hamiltonian can be expressed more conveniently using bosonic creation ($\hat{a}^\dagger$) and annihilation ($\hat{a}$) operators, which represent the addition or removal of a single microwave photon from the resonator. The flux and charge operators are linear combinations of these bosonic operators:

$$\hat{\Phi} = \sqrt{\frac{\hbar Z_r}{2}}(\hat{a} + \hat{a}^\dagger), \hat{Q} = i\sqrt{\frac{\hbar}{2Z_r}}(\hat{a}^\dagger - \hat{a}) \quad (5)$$

where $Z_r = \sqrt{L/C}$ is the characteristic impedance of the resonator. Substituting these into the Hamiltonian yields the familiar form for a quantum harmonic oscillator:

$$\hat{H}_{res} = \hbar\omega_r\left(\hat{a}^\dagger\hat{a} + \frac{1}{2}\right) \quad (6)$$

Here, $\omega_r = 1/\sqrt{LC}$ is the resonant frequency, while $\hat{a}^\dagger$ and $\hat{a}$ are the bosonic creation and annihilation operators for microwave photons within the resonator. The Hamiltonian's eigenvalues, $E_n = \hbar\omega_r(n + 1/2)$, describe a ladder of equally spaced energy levels separated by the single-photon energy $\hbar\omega_r$. The term $\frac{1}{2}\hbar\omega_r$ represents the zero-point energy, a fundamental consequence of the quantization of the electromagnetic field. The operator $\hat{n} = \hat{a}^\dagger\hat{a}$ is the photon number operator, whose integer eigenvalues ($n = 0,1,2,...$) correspond to the number of photons occupying a given state.

When a superconducting qubit (an artificial atom) is coupled to such a resonator, the combined system is described within the powerful framework of *circuit quantum electrodynamics (cQED)*[7,7,86–89]. This framework, inspired by its quantum optics counterpart (cavity QED), treats the qubit as a two-level system interacting with the quantized microwave photon field of the resonator. The total Hamiltonian of the system is the sum of the individual Hamiltonians for the qubit and resonator, plus an interaction term: $\hat{H} = \hat{H}_{res} + \hat{H}_{qubit} + \hat{H}_{int}$. The qubit Hamiltonian, which describes the energy structure of a two-level quantum system, can be derived from first principles. In its general form, expressed in the energy eigenbasis of the ground state $|g\rangle$ and excited state $|e\rangle$, the Hamiltonian is given by: $\hat{H}_{\text{qubit}} = E_g|g\rangle\langle g| + E_e|e\rangle\langle e|$, where $E_g$ and $E_e$ are the energy eigenvalues of the ground and excited states, respectively. What matters physically is the energy difference between these states, which defines the qubit transition frequency $\omega_q$, $E_e - E_g = \hbar\omega_q$. Since the overall energy scale is arbitrary – any constant energy shift results only in an unobservable global phase – we can redefine the zero of energy at the midpoint between $E_g$ and $E_e$. This gives the symmetrically centered energies: $E'_g = -1/2\hbar\omega_q$, $E'_e = +1/2\hbar\omega_q$. Substituting these into the Hamiltonian gives: $\hat{H}_{\text{qubit}} = \frac{\hbar\omega_q}{2}(|e\rangle\langle e| - |g\rangle\langle g|)$. This form can be further simplified using Pauli operators, which naturally describe any two-level system. The Pauli-Z operator $\hat{\sigma}_z$ is conventionally defined such that it has eigenvalue $+1$ for the ground state $|g\rangle$ and $-1$ for the excited state $|e\rangle$, giving: $\hat{\sigma}_z = |g\rangle\langle g| - |e\rangle\langle e|$. Comparing this with the Hamiltonian above, we see that the term in parentheses is equal to $-\hat{\sigma}_z$, leading to the standard form of the qubit Hamiltonian:

$$\hat{H}_{\text{qubit}} = -\frac{1}{2}\hbar\omega_q\hat{\sigma}_z \quad (7)$$

This compact expression captures the essential dynamics of an isolated qubit.

The interaction between the qubit and resonator is fundamentally a dipole coupling, where the qubit's charge or flux dipole interacts with the resonator's electromagnetic field. The general Hamiltonian for this interaction is given by the *quantum Rabi model*:

$$\hat{H}_{int,Rabi} = \hbar g(\hat{a} + \hat{a}^\dagger)\hat{\sigma}_x \quad (8)$$

Here, $g$ is the fundamental coupling strength, a rate determined by the specific physical parameters of the circuit, such as the coupling capacitance and resonator impedance. This Hamiltonian couples the resonator's field amplitude ($\propto \hat{a} + \hat{a}^\dagger$) to the qubit's dipole operator ($\propto \hat{\sigma}_x$). By expanding the Pauli operator $\hat{\sigma}_x = (\hat{\sigma}_+ + \hat{\sigma}_-)$, we can see the four distinct physical processes this interaction describes: $\hat{H}_{int,Rabi} = \hbar g(\hat{a}\hat{\sigma}_+ + \hat{a}^\dagger\hat{\sigma}_- + \hat{a}^\dagger\hat{\sigma}_+ + \hat{a}\hat{\sigma}_-)$. The first two terms, $\hat{a}\hat{\sigma}_+$ (a resonator photon is absorbed to excite the qubit) and $\hat{a}^\dagger\hat{\sigma}_-$ (the qubit de-excites and emits a photon into the resonator), are energy-conserving processes when the qubit and resonator are nearly resonant. The latter two terms, $\hat{a}^\dagger\hat{\sigma}_+$ (a photon is created *and* the qubit is excited) and $\hat{a}\hat{\sigma}_-$ (a photon is absorbed *and* the qubit de-excites), represent processes that violate energy conservation by a large amount, approximately $2\hbar\omega_q$. In most experimental situations, the coupling is in the "strong coupling" but not "ultrastrong coupling" regime, meaning the coupling strength $g$ is much smaller than the resonator and qubit frequencies ($g \ll \omega_r, \omega_q$), while the two are tuned to be nearly resonant ($\omega_r \approx \omega_q$). In this limit, the energy-non-conserving terms oscillate extremely rapidly and their effects average to zero over the characteristic timescale of the interaction, $1/g$. We can therefore safely neglect them. This simplification is known as the Rotating Wave Approximation (RWA)[90].

Applying the RWA and combining the simplified interaction term with the free Hamiltonians for the qubit and resonator yields the celebrated *Jaynes-Cummings (JC) model*[91–93]:

$$\hat{H}_{JC} = \hbar\omega_r \hat{a}^\dagger \hat{a} + \frac{1}{2}\hbar\omega_q \hat{\sigma}_z + \hbar g(\hat{a}^\dagger \hat{\sigma}_- + \hat{a}\hat{\sigma}_+) \quad (9)$$

Here, we have omitted the constant zero-point energy terms for clarity. The operators $\hat{\sigma}_- = |g\rangle\langle e|$ and $\hat{\sigma}_+ = |e\rangle\langle g|$ are the qubit lowering and raising operators, respectively. The term $\hat{a}^\dagger \hat{\sigma}_-$ describes the process where the qubit de-excites from $|e\rangle$ to $|g\rangle$ while creating a photon in the resonator, and $\hat{a}\hat{\sigma}_+$ describes the time-reversed process. The parameter $g$ now clearly represents the coherent coupling strength, quantifying the rate at which a single quantum of energy is resonantly exchanged between the qubit and the resonator. The JC model is one of the most fundamental and important models in quantum optics and has been realized with unprecedented control and fidelity in circuit QED systems, forming the basis of most quantum gate operations and measurement schemes.

### 2.4 From Single Resonators to Metamaterials: The Jaynes-Cummings-Hubbard Model

The true power of metamaterials comes from creating periodic arrays of these fundamental circuit elements. An array of coupled resonators, each containing a qubit, can no longer be described by the simple JC model. Instead, it forms a quantum many-body system whose behavior is captured by the *Jaynes-Cummings-Hubbard (JCH) model*[94–96]. The JCH model extends the JC model to a lattice, adding a term that describes the hopping of photons between adjacent resonator sites:

$$\hat{H}_{JCH} = \sum_{j=1}^{N}\left(\hbar\omega_r \hat{a}_j^\dagger \hat{a}_j + \frac{1}{2}\hbar\omega_q \hat{\sigma}_{z,j} + \hbar g_j(\hat{a}_j^\dagger \hat{\sigma}_{-,j} + \hat{a}_j \hat{\sigma}_{+,j})\right) - \sum_{\langle i,j \rangle} \kappa_{ij}(\hat{a}_i^\dagger \hat{a}_j + \hat{a}_j^\dagger \hat{a}_i) \quad (10)$$

In this Hamiltonian, the first sum runs over all $N$ sites of the lattice, with each site described by its own JC-like Hamiltonian; the second sum runs over pairs of neighboring sites $\langle i,j \rangle$; $\kappa_{ij}$ is the photon tunneling or hopping rate between resonator $i$ and resonator $j$. This parameter is determined by the physical coupling (e.g., capacitive or inductive) between the resonators.

The JCH model describes a rich landscape of quantum phenomena. The elementary excitations of this system are polaritons – hybrid quasiparticles of light and matter. The competition between the on-site interaction strength $g$ and the inter-site hopping rate $\kappa$ can lead to *quantum phase transitions*[95,97,98]. For strong on-site interactions ($g \gg \kappa$), the system enters a *Mott insulator phase*, where polaritons are localized at individual sites. Conversely, for strong hopping ($\kappa \gg g$), the polaritons delocalize and form a *superfluid phase* across the entire lattice[95,99–104]. This model provides the theoretical foundation for using superconducting metamaterials as quantum simulators to study complex many-body physics and to realize novel states of matter, such as topological phases of light. The ability to start from a classical circuit diagram and arrive at a predictive quantum many-body Hamiltonian like the JCH model is a testament to the power and versatility of the cQED framework.

### 3. Electromagnetic Environment Engineering: Design and Dispersion

The remarkable success of superconducting circuits for quantum information processing has been built upon the ability to precisely control the electromagnetic environment in which artificial atoms (qubits) are embedded. This control determines fundamental properties such as qubit lifetimes, interaction strengths, and readout fidelity. Initially, this environment was provided by standard microwave engineering components, such as distributed transmission line resonators. However, the quest for scalability and enhanced functionality has driven a paradigm shift towards using metamaterial concepts to create bespoke electromagnetic environments with properties not found in conventional systems. A metamaterial, in this context, is an artificial structure composed of a periodic arrangement of sub-wavelength resonant elements, or "meta-atoms," whose collective behavior, rather than the intrinsic properties of their constituent materials, dictates the overall electromagnetic response. This approach unlocks the ability to engineer the dispersion relation – the fundamental relationship between frequency ($\omega$) and wave number ($k$) – providing an unprecedented level of control over the density of photonic states, mode structure, and impedance[36,40].

## 3.1 From Transmission Lines to Lumped-Element Resonators: A Miniaturization Revolution

The foundational component for coupling and reading out qubits in early circuit QED experiments was the distributed resonator, most commonly realized as a coplanar waveguide (CPW) resonator[7,86,89,105,106]. A CPW consists of a central signal line separated from two flanking ground planes on a dielectric substrate. A resonator is formed by taking a finite length of this transmission line and terminating it in a way that confines electromagnetic waves, for instance, by leaving the ends open (capacitive coupling) or shorting them to ground (inductive coupling). These structures function analogously to a musical instrument's string; they support standing waves only at specific frequencies where the resonator's length, $L$, is an integer multiple of a half-wavelength of the guided microwave signal. For a half-wave ($\lambda/2$) resonator, the resonant frequencies are given by: $\omega_n = n \frac{\pi v_p}{L}, \quad n = 1,2,3,...$, where $v_p = c/\sqrt{\epsilon_{\text{eff}}}$ is the phase velocity of the wave, determined by the speed of light $c$ and the effective dielectric constant $\epsilon_{\text{eff}}$ of the substrate medium (e.g., silicon or sapphire)[87,105].

This distributed approach, while successful for proof-of-principle experiments, presents two severe limitations for building large-scale quantum processors. The first one is, the large physical footprint: the resonant frequency is directly tied to the physical length. For a typical qubit operating frequency of $f = 6$ GHz on a sapphire substrate with $\epsilon_{\text{eff}} \approx 10$, the guided wavelength is $\lambda_g = v_p/f \approx (c/\sqrt{10})/(6 \times 10^9 \text{ Hz}) \approx 1.58$ cm. A half-wave resonator would thus be nearly 8 mm long[86]. This macroscopic size consumes an untenable amount of chip real estate, making it difficult to densely pack the thousands of resonators needed for a scalable architecture. The second limitation is the sparse, inflexible spectrum: the resonant modes are harmonically spaced, with $\omega_2 = 2\omega_1$, $\omega_3 = 3\omega_1$, and so on. This sparse spectrum offers little flexibility. Placing multiple qubits at distinct frequencies for individual addressing becomes challenging, and the large frequency gaps between modes are ill-suited for creating the dense multi-mode structures needed for advanced applications like quantum simulation or serving as a multi-channel quantum bus[107].

The solution to these problems lies in transitioning from distributed elements to lumped elements. An electrical component is considered lumped if its physical dimensions, $d$, are much smaller than the wavelength of the signal it carries, i.e., $d \ll \lambda_g/20$. While a CPW resonator's behavior is distributed along its entire length, components like small spiral inductors and interdigitated capacitors can be fabricated on the micron scale, easily satisfying the lumped-element condition. By combining a lumped inductor ($L$) and capacitor ($C$), one can create a simple LC harmonic oscillator with a resonant frequency given by $\omega_r = 1/\sqrt{LC}$. This fundamental shift has profound consequences. The resonator's frequency is no longer determined by its physical length but by the designed values of $L$ and $C$. This decouples the operational frequency from the device footprint, enabling extreme miniaturization and freeing up chip space. More importantly, it provides a new degree of freedom for engineering the circuit's properties. A resonator is no longer just a piece of wire of a certain length; it is a designed circuit with tunable parameters. This conceptual leap is the gateway to metamaterial engineering. By arranging these compact LC oscillators into a periodic one-dimensional or two-dimensional array, one constructs a metamaterial transmission line or surface. The properties of this new, artificial medium are not determined by a bulk material constant like $\epsilon_{\text{eff}}$, but by the engineered parameters of its constituent meta-atoms ($L$ and $C$) and their coupling. This process of sub-wavelength engineering – controlling macroscopic wave propagation by designing the microscopic structure of the medium – is the core principle that metamaterials bring to quantum circuits.

## 3.2 Left-Handed Metamaterials: Unconventional Dispersion and Dense Mode Spectra

A particularly powerful and transformative example of dispersion engineering is the creation of a "left-handed" (LH) metamaterial[23,108]. First theorized by Veselago in the context of continuous media[109], these materials exhibit simultaneously negative effective permittivity ($\epsilon_{eff} < 0$) and permeability ($\mu_{eff} < 0$)[38,110–114]. This leads to a wave vector **k**, electric field **E**, and magnetic field **H** that form a left-handed triad, in contrast to the right-handed triad found in conventional media[115]. The most striking consequence is that the group velocity ($v_g = d\omega/dk$), which represents the direction of energy flow, and the phase velocity ($v_p = \omega/k$) are antiparallel.

In a circuit context, this unconventional behavior is achieved by constructing a periodic lattice that is the dual of a conventional right-handed (RH) transmission line. Whereas an RH line can be modeled as a chain of series inductors ($L_R$) and shunt capacitors ($C_R$), a pure LH transmission line consists of series capacitors ($C_L$) and shunt inductors ($L_L$)[116–119]. Using Bloch's theorem for periodic structures, one can derive the dispersion relation for an ideal, lossless LH transmission line composed of unit cells of length $\Delta x$. A simplified but widely used form that captures the essential physics is[116,120,121]:

$$\omega(k) = \frac{1}{2\sqrt{L_L C_L}|\sin(k\Delta x/2)|} \quad (11)$$

This equation reveals the hallmark of LH dispersion: as the wave number $k$ increases from zero, the frequency $\omega$ *decreases* from an upper cutoff, which is precisely the opposite behavior of conventional RH lines where frequency increases with wave number[122,123]. In any practical implementation, parasitic effects from the physical layout inevitably introduce some right-handed inductance and capacitance. This leads to a more general and realistic model known as the composite right/left-handed (CRLH) transmission line. The dispersion relation for a lossless CRLH line is given by: $\omega^2(k) = \frac{1}{L_R C_L} + \frac{1}{L_L C_R} - \frac{2}{\sqrt{L_R C_R L_L C_L}} \cos(k\Delta x)$. This more complete model shows a passband between a lower and upper cutoff frequency, with the left-handed character dominating at lower frequencies and the right-handed character at higher frequencies.

A profound consequence of this inverted dispersion, when implemented in a finite-length resonator, is the emergence of a dense spectrum of orthogonal microwave modes at frequencies *above* a low-frequency bandgap where propagation is forbidden. This is ideal for cQED architectures. It allows a large number of addressable resonator modes to be packed into the typical 4–8 GHz operating window of superconducting qubits, all within a compact physical footprint. This transforms the resonator from a simple single-mode component into a rich, multimode quantum bus, providing an essential resource for applications like multi-qubit coupling, analog quantum simulation of many-body systems, and hardware-efficient quantum memories. For example, by forming a left-handed metamaterial into a ring, one can mediate tunable, long-range entangling interactions between multiple qubits coupled at different points around the ring[23,52].

### 3.3 Engineering the Resonator Impedance and Qubit Coupling

The strength of the interaction between a qubit and a resonator is a critical parameter that dictates the speed of quantum operations and readout[7,89,106,107,124–126]. For capacitive coupling between a transmon qubit and a resonator, this strength, $g$, is proportional to the zero-point voltage fluctuations, $V_{rms}$, of the resonator mode, which are in turn determined by the resonator's characteristic impedance, $Z_r$. The coupling strength can be expressed as $g \propto \beta V_{rms} \propto \frac{C_c}{C_\Sigma} \sqrt{\frac{\hbar \omega_r Z_r}{2}}$, where $C_c$ is the coupling capacitance, $C_\Sigma$ is the total capacitance of the qubit, $\beta = C_c/C_\Sigma$ is the coupling ratio, $\omega_r$ is the resonator frequency, and the characteristic impedance is defined by the per-unit-cell inductance $L$ and capacitance $C$ as $Z_r = \sqrt{L/C}$ [7,127]. This crucial relationship shows that achieving stronger qubit-resonator coupling requires designing resonators with higher impedance[86,87,128–132].

Metamaterial engineering provides a direct and powerful route to achieving exceptionally high characteristic impedance[133]. Conventional CPW resonators are typically designed to have an impedance of 50 Ω to match standard microwave equipment[123]. However, by constructing resonators from metamaterial unit cells with very high inductance, characteristic impedances of several kilo-ohms can be realized[134–139]. This high inductance can be achieved geometrically, for example, by using long, thin nanowires or tightly wound spiral inductors, or by exploiting the high kinetic inductance of materials like NbN or granular aluminum[67,70,137,138,140–148]. An even more powerful approach is to use arrays of Josephson junctions, which act as tunable, high-quality superinductors[149–153].

This capability allows circuits to be pushed from the standard strong coupling regime ($g$ exceeds the qubit and resonator decay rates, $\kappa, \gamma$) into the ultrastrong coupling (USC) regime, where the coupling strength becomes a significant fraction of the resonator frequency ($g/\omega_r \gtrsim 0.1$), and even into the deep-strong coupling (DSC) regime, where $g/\omega_r \geq 1$ [86,87,154–165]. In these regimes, the widely used rotating wave approximation (RWA), which assumes $g \ll \omega_r$ and neglects fast-oscillating terms, breaks down. The system can no longer be described by the Jaynes-Cummings Hamiltonian Eq.(9) and must instead be described by the full quantum Rabi Hamiltonian: $\hat{H}_{Rabi} = \hbar\omega_r\left(\hat{a}^\dagger\hat{a} + \frac{1}{2}\right) + \frac{1}{2}\hbar\omega_q\hat{\sigma}_z + \hbar g(\hat{a}^\dagger + \hat{a})\hat{\sigma}_x$. The counter-rotating terms $(\hat{a}^\dagger\hat{\sigma}_+ + \hat{a}\hat{\sigma}_-)$, which are neglected in the Jaynes-Cummings model, become significant and lead to qualitatively new physics. The ground state of the system is no longer a simple vacuum state but becomes a two-mode squeezed vacuum, containing a finite population of virtual photons and exhibiting entanglement between the qubit and resonator[154,161,166–172]. The ability to engineer both the dispersion and the impedance of the electromagnetic environment is a defining feature of the metamaterial approach, enabling the exploration of novel physical regimes and providing a crucial resource for advanced quantum information processing.

## 4. Materials, Fabrication, and Coherence Limits

The promise of metamaterial resonators for quantum – technologies – offering engineered dispersion, high impedance, and compact footprints – hinges on the ability to fabricate these complex, structured devices while maintaining the exceptionally high quantum coherence required for computation[36,39,41,42,48,52,80,173–175]. This presents a fundamental tension: metamaterial functionality arises from engineered geometric complexity, which inherently increases the number of surfaces and interfaces, the very locations where performance-limiting defects reside[11,17,176–187]. Overcoming this challenge requires a holistic approach that integrates materials science, advanced nanofabrication, and novel device architectures. This section bridges the gap between the theoretical promise of quantum metamaterials and the practical realities of their implementation, surveying the material platforms, dominant decoherence channels, and state-of-the-art fabrication strategies that define the field.

### 4.1 Material Platforms: From Workhorses to Hybrids

The choice of superconducting material is a primary determinant of device performance, influencing not only the intrinsic loss but also key design parameters like kinetic inductance and compatibility with fabrication processes.

***The Workhorses: Niobium and Aluminum.*** Historically, Aluminum (Al) and Niobium (Nb) have been the workhorses of the field, largely due to their robust, conventional superconducting properties and their compatibility with well-established microfabrication techniques derived from the semiconductor industry[54,188–198]. Aluminum, with its low melting point and simple oxide ($Al_2O_3$), is particularly well-suited for fabricating Josephson junctions using shadow evaporation techniques. Niobium, with its higher critical temperature ($T_c \approx 9.2$ K), offers greater thermal robustness. However, the native oxides of both materials, particularly niobium, have been identified as significant sources of dielectric loss, a problem that has spurred intensive research into surface treatments and alternative materials.

***High-Kinetic-Inductance Materials.*** The quest for higher impedance and greater compactness has driven the adoption of materials with high kinetic inductance ($L_k$). As described by the formula $L_k = \hbar R_s/(\pi\Delta)$, materials with high normal-state sheet resistance ($R_s$) act as "superinductors" in their superconducting state. This category includes disordered superconductors such as Niobium Nitride (NbN), Titanium Nitride (TiN), and granular aluminum (grAl)[8,67,67,70,138,199–206]. These materials allow for the creation of extremely compact, high-impedance resonators and are essential for fluxonium qubits, which rely on a large shunting inductance. A systematic comparison of NbN and grAl, for instance, revealed that while grAl exhibits a much higher nonlinear Kerr coefficient (advantageous for certain quantum operations), NbN offers superior resilience to magnetic fields, making it more suitable for hybrid quantum systems that involve magnetic components. The exploration continues with novel candidates like aluminum nitride (AlNx), which is expected to be a promising superinductor[207,208].

***Breakthroughs with Tantalum.*** A recent and significant breakthrough has been the adoption of Tantalum (Ta) as a primary material for superconducting qubits[173,209–216]. Transmon qubits fabricated with tantalum have demonstrated record-breaking coherence times exceeding 0.3 ms, with some reports reaching as high as 0.6 ms[210,211]. This dramatic improvement is attributed to the material properties of tantalum's native oxide, $Ta_2O_5$. Density functional theory calculations suggest that $Ta_2O_5$ forms a smoother surface with fewer structural defects and dangling atoms compared to niobium oxide. These defects are believed to be the microscopic origin of two-level systems (TLSs), the dominant source of decoherence. The lower density of TLSs in tantalum oxide directly translates to lower dielectric loss and, consequently, longer qubit lifetimes.

***The Frontier: Hybrid Quantum Systems.*** A burgeoning frontier is the development of hybrid quantum systems, where superconducting circuits are integrated with other material platforms to explore novel physics or create new functionalities[217–226]. Of particular interest are two-dimensional van der Waals (vdW) materials, which span a vast range of properties including magnetism, topology, and unconventional superconductivity[69,227,228]. Integrating vdW flakes – such as the high-temperature superconductor $Bi_2Sr_2CaCu_2O_{8+x}$ (BSCCO) or magic-angle twisted bilayer graphene (MATBG) – with a conventional niobium resonator allows for probing the microscopic properties of the vdW material using the high sensitivity of the superconducting circuit[229–234]. However, creating a pristine, high-coherence interface between such disparate materials presents a significant fabrication challenge, requiring advanced techniques like cryogenic stacking and the use of amorphous silicon capping layers to prevent oxidation and degradation of the superconducting film[234–238]. Despite these challenges, hybrid BSCCO-niobium resonators have been demonstrated with high quality factors ($Q \approx 3 \times 10^4$), opening a path for exploring exotic material physics within a cQED framework[235].

### 4.2 Dominant Loss Channels: The War on Decoherence

Despite the absence of DC resistance, superconducting quantum circuits are subject to several microwave-frequency loss mechanisms that limit coherence times and gate fidelities. The total quality factor of a resonator, $Q_L$, is limited by both internal (intrinsic) losses of the device, quantified by $Q_i$, and external losses due to coupling to the measurement circuitry, $Q_c$. The relationship is given by:

$$\frac{1}{Q_L} = \frac{1}{Q_i} + \frac{1}{Q_c} \quad (12)$$

While $Q_c$ is an engineered parameter, minimizing internal loss (maximizing $Q_i$) is a central goal of materials research.

***Two-Level Systems (TLSs): The Ubiquitous Adversary.*** The dominant source of energy loss in most high-coherence superconducting qubits and resonators, especially at the single-photon power levels relevant for quantum computation, is coupling to spurious two-level systems (TLSs)[20,185,186,239–247]. TLSs are microscopic defects, such as tunneling atoms or atomic configurations, that reside primarily in amorphous dielectric materials. They are found ubiquitously at the critical interfaces of a device: the metal-air (MA) interface, the metal-substrate (MS) interface, and the substrate-air (SA) interface. The amorphous native oxides that form on superconductors and the tunnel barrier of the Josephson junction itself are particularly dense with TLSs.

While a comprehensive microscopic explanation remains contested, the standard tunneling model treats a TLS as a particle in a local anharmonic double-well potential[248,249]. This gives rise to a broad distribution of energy splittings. TLSs whose energy splitting is resonant with the qubit frequency can absorb a microwave photon, leading to energy relaxation (a limit on $T_1$). The collective behavior of the TLS bath also creates fluctuations in the local dielectric environment, causing the qubit's frequency to jitter and leading to pure dephasing (a limit on $T_\phi$). The total decoherence rate $1/T_2$ is given by the sum of relaxation and dephasing contributions:

$$\frac{1}{T_2} = \frac{1}{2T_1} + \frac{1}{T_\phi} \quad (13)$$

A key signature of TLS-induced loss is its dependence on microwave power. At high powers, the TLSs become saturated and can no longer absorb energy, causing the loss to decrease. This behavior is captured by the following formula for the TLS-limited internal quality factor[182,250–252]:

$$\frac{1}{Q_{i,TLS}(P)} = \frac{F\tan\delta_{TLS}}{\sqrt{1 + (P/P_c)^\beta}}, \quad (14)$$

where $F$ is a geometric filling factor describing the fraction of electric field energy stored in the TLS-containing dielectric, $\tan\delta_{TLS}$ is the intrinsic loss tangent of the material, $P$ is the microwave power, $P_c$ is a characteristic saturation power, and $\beta$ is a fitting exponent typically between 0.2 and 0.5. In addition to causing dissipation, the TLS bath also induces a temperature-dependent frequency shift in the resonator, $\delta f$. This effect provides another spectroscopic tool for characterizing the TLS environment.

***Other Loss Channels.*** While TLSs are often dominant, other mechanisms contribute to decoherence. *Non-equilibrium quasiparticles* (unpaired electrons) can be generated by stray infrared radiation or high-energy particles like cosmic rays breaking Cooper pairs, providing another channel for dissipation[253–255]. *Trapped magnetic flux vortices*, which can be pinned at material defects, also dissipate energy as they move under the influence of microwave currents[256–260]. Finally, *surface contamination*, including chemical residues from fabrication and stray paramagnetic spins, can also couple to the qubit and induce loss[11,261–264].

### 4.3 Fabrication Strategies for High-Coherence Devices

Mitigating the aforementioned loss channels requires sophisticated fabrication strategies that go beyond simple circuit patterning and involve a co-design of the device's physical and electromagnetic properties.

***Surface and Interface Engineering.*** Since interfaces are a primary source of TLS loss, their treatment is paramount[213,265–268]. This begins with substrate cleaning, where techniques like hydrofluoric acid (HF) treatment and high-temperature vacuum annealing are used to remove native oxides and contaminants from silicon wafers before superconductor deposition. Post-deposition processing is also critical. Argon ion milling, a common step to clean surfaces before depositing a subsequent layer, can itself introduce subsurface damage and create more TLS defects, particularly in niobium. However, it has been shown that a subsequent gentle dry etch can fully recover these losses, offering a path to high-quality multilayer circuits[269]. To prevent the re-formation of lossy native oxides after fabrication, protective capping or passivation layers are employed. Encapsulating tantalum with a few nanometers of a noble metal like gold (Au) or a gold-palladium (AuPd) alloy has been shown to completely suppress the formation of tantalum oxide, effectively removing a major source of loss[210].

***Architectural Solutions: Suspending Circuits.*** A more radical approach to eliminating interface loss is to physically remove the interface itself. Recent work has demonstrated a versatile technique for selectively etching the substrate from underneath key circuit components, such as Josephson junction arrays or sections of a resonator[270–273]. This suspends the component in vacuum, drastically reducing the electric field participation in the lossy substrate dielectric and eliminating the metal-substrate interface entirely. This method systematically reduces stray capacitance and increases inductance without degrading the quality of the superconducting elements, providing a powerful tool for isolating and mitigating substrate-related loss mechanisms.

***Scalability: Fabrication Tolerances and 3D Integration.*** A major challenge in scaling up any quantum technology based on artificial atoms is fabrication-induced inhomogeneity. Unlike natural atoms, which are identical, variations in lithography and etching processes lead to a spread in the parameters of superconducting qubits, most notably their resonant frequencies[194,246,273,274]. For metamaterials that rely on the collective, coherent response of many meta-atoms, this disorder can be detrimental. To address this, there is a strong push to move from laboratory-scale fabrication to industry-standard, 300 mm CMOS pilot lines[194]. These advanced facilities offer superior process control, enabling higher yield and uniformity across large wafers, which is a prerequisite for building complex quantum processors.

As planar circuits become increasingly dense, *3D integration* via techniques like flip-chip bonding offers a promising path to scalability[275–278]. This allows for a modular design, separating the sensitive qubit layer from the potentially noisy control and readout electronics on another chip[279]. However, this introduces new challenges, including the need for high-precision alignment and the introduction of new materials (e.g., indium bumps, dielectric bonding layers) that can themselves be sources of loss and decoherence. Mitigating crosstalk between layers and managing the thermal budget are also critical concerns that must be addressed to realize the full potential of 3D quantum architectures.

## 5. State-of-the-Art Performance and Record Values

The rapid progress in designing, fabricating, and understanding superconducting metamaterial resonators is reflected in the continuous improvement of key performance benchmarks. The theoretical promise of engineered quantum environments is ultimately validated by tangible, record-setting experimental results. These metrics – resonator quality factor, qubit coherence time, and light-matter coupling strength – are the yardsticks by which the field measures its advance toward fault-tolerant quantum computation. They are often deeply intertwined; for instance, a high-quality resonator is a prerequisite for achieving long qubit coherence, and a high-impedance metamaterial is necessary to reach unprecedented coupling strengths. This section will survey the state of the art in these critical areas, presenting the record values that define the frontiers of what is possible with superconducting and cryogenic quantum technologies.

### 5.1 Pushing the Limits of Quality Factors in Superconducting Resonators

The quality factor ($Q$) of a resonator is a dimensionless parameter that quantifies its ability to store energy relative to the rate at which that energy is dissipated[123]. It is fundamentally defined as the ratio of the energy stored in the resonator to the energy lost per radian of the oscillation cycle. In practical terms, it is often measured as the ratio of the resonator's center frequency, $f_r$, to its resonance linewidth or full-width at half-maximum (FWHM), $\Delta f$: $Q = \frac{f_r}{\Delta f}$. A high $Q$-factor signifies low energy loss and, consequently, a long lifetime for the photons stored within the resonator. This is crucial for numerous quantum applications, including the development of long-lived quantum memories, the implementation of high-fidelity qubit readout schemes, and the construction of sensitive detectors for single photons or other weak signals[11]. The experimentally measured or "loaded" quality factor, $Q_L$, is always limited by two distinct loss channels: internal (or intrinsic) losses inherent to the device itself, quantified by the internal quality factor $Q_i$, and external losses due to the intentional coupling of the resonator to the external measurement circuitry, quantified by the external or coupling quality factor $Q_c$. The total loss is the sum of the individual loss rates, leading to Eq.(12). By designing coupling to be very weak ($Q_c \gg Q_i$), the loaded quality factor becomes approximately equal to the internal one ($Q_L \approx Q_i$), allowing for a direct measurement of the intrinsic material and fabrication quality of the device. The pursuit of higher $Q_i$ has been a central theme in the development of superconducting circuits, leading to remarkable achievements across different resonator platforms.

***Three-Dimensional (3D) Cavities***: The highest quality factors have been consistently demonstrated in 3D microwave cavities[11,261,280,281]. These are typically macroscopic structures, often machined from bulk, high-purity niobium, where the electromagnetic field is confined within a three-dimensional volume[263,282–284]. The key advantage of this architecture is the minimization of surface participation; the vast majority of the electromagnetic energy is stored in the vacuum of the cavity, with only a tiny fraction interacting with the lossy surfaces of the cavity walls. Through decades of materials development and the refinement of surface treatment techniques – such as electropolishing and high-temperature baking to remove surface defects and contaminants – these 3D cavities have achieved astonishing internal quality factors exceeding $10^9$ and, in some cases, approaching $10^{11}$. Such high $Q_i$ values correspond to photon lifetimes on the order of seconds, making them exceptional platforms for quantum memories and for housing highly coherent qubits. Recent work continues to push these boundaries, with demonstrations of $Q_{int} \gtrsim 1.4 \times 10^9$ in the single-photon regime, representing a significant improvement over previous benchmarks.

*Planar Resonators:* While 3D cavities offer the highest performance, planar resonators fabricated using lithographic techniques are far more amenable to the large-scale integration required for quantum processors. In these devices, achieving high $Q_i$ is more challenging because a much larger fraction of the electric field interacts with potentially lossy surfaces and interfaces. Nevertheless, tremendous progress has been made through careful materials selection and interface engineering. By fabricating aluminum resonators on ultra-low-loss sapphire substrates and employing meticulous substrate preparation and cleaning protocols, internal quality factors approaching and exceeding 2 million have been demonstrated at single-photon power levels[18,181]. More recently, geometric design has emerged as a powerful tool for loss reduction. Archimedean spiral resonators (ASRs), for example, have been shown to have lower electric field participation at surfaces compared to standard coplanar waveguides, enabling the realization of intrinsic quality factors approaching 10 million ($Q_i = (9.6 \pm 1.5) \times 10^6$) in titanium nitride (TiN) devices[285].

**Cryogenic and Metamaterial Resonators:** The quest for high $Q$ extends beyond superconducting circuits. Cryogenic resonators made from silicon nitride have achieved $Q$-factors up to $1.2 \times 10^8$ at 14 mK, demonstrating the potential of low-loss dielectric materials[286]. Similarly, microfabricated bulk acoustic wave resonators have reached $Q$-factors of $2.8 \times 10^7$ at cryogenic temperatures[287]. Within the realm of metamaterials, which often involve more complex and loss-prone geometries, achieving high quality factors is a significant challenge[288–290]. However, novel design principles offer unique pathways to suppress loss. For instance, metamaterials designed to support non-radiating anapole modes can, in principle, achieve extremely high $Q$-factors by eliminating radiative losses through destructive interference of their constituent dipole moments[291–296]. While experimental demonstrations in superconducting circuits are still emerging, theoretical work on terahertz metasurfaces suggests that such anapole-based approaches could yield $Q$-factors exceeding $10^4$[297].

### 5.2 Record Coherence Times in Metamaterial-Coupled Qubits

Ultimately, the performance of a quantum processor is determined by the coherence of its fundamental building blocks, the qubits. Qubit coherence is characterized by two primary timescales: the energy relaxation time, $T_1$, which measures how long a qubit remains in its excited state before decaying to the ground state, and the dephasing time, $T_2$, which measures how long the qubit can maintain a definite phase relationship in a quantum superposition. The total decoherence rate is given by Eq.(13). Over the past two decades, a combination of improved qubit design, materials science, and fabrication techniques has led to a staggering, multi-order-of-magnitude improvement in these coherence times.

***Transmon Qubits:*** The transmon qubit, a design that is intrinsically protected from charge noise, has been a workhorse of the field. For years, its coherence times hovered in the tens to low hundreds of microseconds[211,213]. A major breakthrough came with the strategic replacement of niobium with tantalum in the fabrication of the qubit's large capacitor pads. This material substitution led to a dramatic suppression of decoherence, with reports demonstrating both $T_1$ and $T_2$ times exceeding 0.5 milliseconds[173,213,215,298–300]. This improvement is attributed to the superior properties of tantalum's native oxide, which appears to host fewer TLS defects[215,301]. Further optimization of materials and fabrication, including the use of noble metal capping layers to prevent oxidation entirely, has pushed these numbers even higher. Recent work has reported reproducible transmon lifetimes with average values of 0.3 ms and maximum values as high as 0.6 ms[210]. In a landmark 2024 result, researchers characterized a transmon with a median $T_1$ of ~500 μs and a maximum echo-dephasing time ($T_{2,echo}$) of ~1 ms, demonstrating that there is no fundamental barrier to achieving millisecond-scale coherence in transmon qubits[302].

***Fluxonium Qubits:*** The fluxonium qubit, a more complex circuit design featuring a high-inductance shunt that provides even greater intrinsic protection against environmental noise, has recently emerged as a top contender for high-coherence applications[298,303–306]. By operating at lower frequencies and carefully optimizing circuit parameters, researchers have systematically pushed the boundaries of fluxonium coherence. In 2023, a landmark experiment demonstrated a fluxonium qubit with a relaxation time of ~1.4 ms, a tenfold improvement over previous records for this qubit type[306]. This milestone places fluxonium on par with, or even ahead of, the best transmons and makes it a highly attractive platform for future quantum processors.

***Metamaterial-Enhanced Coherence:*** Metamaterials offer a direct and powerful route to enhancing these already impressive coherence times by engineering the electromagnetic vacuum itself. A qubit's spontaneous emission rate is governed by Fermi's golden rule, which states that the decay rate is proportional to the density of available photonic states at the qubit's transition frequency. This phenomenon is known as the Purcell effect[93,307,308]. By designing a metamaterial with a photonic bandgap – a range of frequencies where propagation is forbidden – one can dramatically reduce the density of states available for the qubit to decay into. Tuning the qubit's frequency into this bandgap effectively "hides" it from the vacuum, suppressing its spontaneous emission and thereby increasing its $T_1$ lifetime[41,42,309]. This has been experimentally demonstrated by coupling a transmon qubit to a superconducting metamaterial waveguide. When the qubit was tuned to the edge of the metamaterial's bandgap, where the group velocity of light approaches zero and the density of states is highly modified, a remarkable 24-fold enhancement of its radiative lifetime was observed[41]. This powerful demonstration shows that metamaterials can be used not just as passive circuit elements, but as active tools for quantum state protection.

## 5.3 Achieving Strong, Ultrastrong, and Deep-Strong Coupling Regimes

The speed and fidelity of two-qubit gates and readout protocols are governed by the qubit-resonator coupling strength, $g$. The standard paradigm of circuit QED operates in the *strong coupling regime*, where the coherent coupling rate $g$ exceeds the decay rates of both the qubit ($\gamma$) and the resonator ($\kappa$), i.e., $g \gg \kappa, \gamma$. This condition allows for the coherent exchange of a single quantum of energy between the qubit and the resonator to occur many times before decoherence destroys the quantum state[86,89,92,163,310].

Metamaterial resonators, with their unique ability to achieve very high characteristic impedances, provide a platform to explore even more extreme interaction regimes. By constructing resonators from elements with very high inductance, such as arrays of Josephson junctions or high-kinetic-inductance nanowires, impedances $Z_r = \sqrt{L/C}$ well over the resistance quantum ($R_Q = h/e^2 \approx 25.8$ kΩ) can be realized[128,311]. Since the coupling strength scales as $g \propto \sqrt{Z_r}$, these high-impedance resonators enable access to the *ultrastrong coupling (USC)* regime, where $g$ becomes a significant fraction of the resonator frequency $\omega_r$ (typically defined as $g/\omega_r \gtrsim 0.1$), and even the *deep-strong coupling (DSC)* regime, where the coupling strength meets or exceeds the resonator frequency ($g/\omega_r \geq 1$)[124,129,158,312,313]. Experimentally, achieving these regimes in superconducting circuits is an active and challenging frontier. Coupling strengths ($g/2\pi$) exceeding 1 GHz have been realized using high-impedance resonators made from Josephson junction arrays[311] and lumped-element resonator with impedance exceeding 12 kΩ based on a high-kinetic-inductance NbTiN thin film[314]. While reaching the DSC regime in a circuit QED architecture remains an ongoing goal, other quantum simulation platforms have provided a glimpse of the physics involved. In a landmark experiment using trapped cold atoms to simulate the quantum Rabi model, a normalized coupling ratio of $g/\omega \approx 6.5$ was demonstrated, placing the system firmly in the DSC regime and setting a remarkable benchmark for circuit-based platforms to pursue[315,316]. The ability of metamaterials to engineer arbitrarily high impedance provides a clear path forward for realizing these exotic coupling regimes and exploring the rich many-body physics they contain[130,131,140,317–319].

The following table summarizes some of the key record performance metrics that define the state of the art in the field.

**Table 1:** Key Record Performance Metrics.

| Platform/Device Type | Key Parameter | Record Value | Material System | Key Innovation/Concept | Refs. |
|---|---|---|---|---|---|
| 3D Superconducting Cavity | Internal Q-factor ($Q_i$) | $> 1.4 \times 10^9$ | Bulk Niobium | Material purity, advanced surface treatments | 282,320 |
| Planar Spiral Resonator | Internal Q-factor ($Q_i$) | $\approx 9.6 \times 10^6$ | TiN on Silicon | Geometric optimization (spiral vs. CPW) to reduce surface loss participation | 321 |
| Transmon Qubit | Coherence Time ($T_{2,echo}$) | $> 1.0$ ms | Tantalum on Sapphire | Material substitution (Ta for Nb) to reduce TLS loss from surface oxides | 322 |
| Fluxonium Qubit | Coherence Time | $> 1.4$ ms | Al/AlOx/Al on | Optimized circuit parameters, | 306 |

| | ($T_1$) | | Sapphire | operation at low frequency | |
| --- | --- | --- | --- | --- | --- |
| Metamaterial-Coupled Transmon | Lifetime Enhancement | 24x | Nb on Silicon | Purcell inhibition by engineering a photonic bandgap | 41 |
| High-Impedance Resonator | Coupling Strength ($g/2\pi$) | up to 155 MHz | SQUID Array / GaAs | High impedance from JJ array to enhance vacuum fluctuations | 129 |
| Cold Atom Quantum Simulator | Coupling Ratio ($g/\omega$) | ≈ 6.5 | Trapped Rubidium Atoms | Simulation of the quantum Rabi model in the deep-strong coupling regime | 315,316 |

## 5.4 Quantum Simulators for Exploring Many-Body Physics

Recent advances in the fabrication and control of multi-qubit processors have elevated superconducting quantum metamaterials from theoretical concepts to powerful experimental platforms. Beyond their potential for building more robust quantum computers, these systems serve as versatile, large-scale analog quantum simulators for exploring many-body physics. Two distinct, state-of-the-art architectures, illustrated in Fig. 1, highlight this dual promise. The first uses a large ensemble of qubits coupled to a single cavity mode to study the fundamental interplay between collective coherence and disorder, while the second uses a photonic-bandgap waveguide to engineer a many-body Hamiltonian with tunable, long-range interactions.

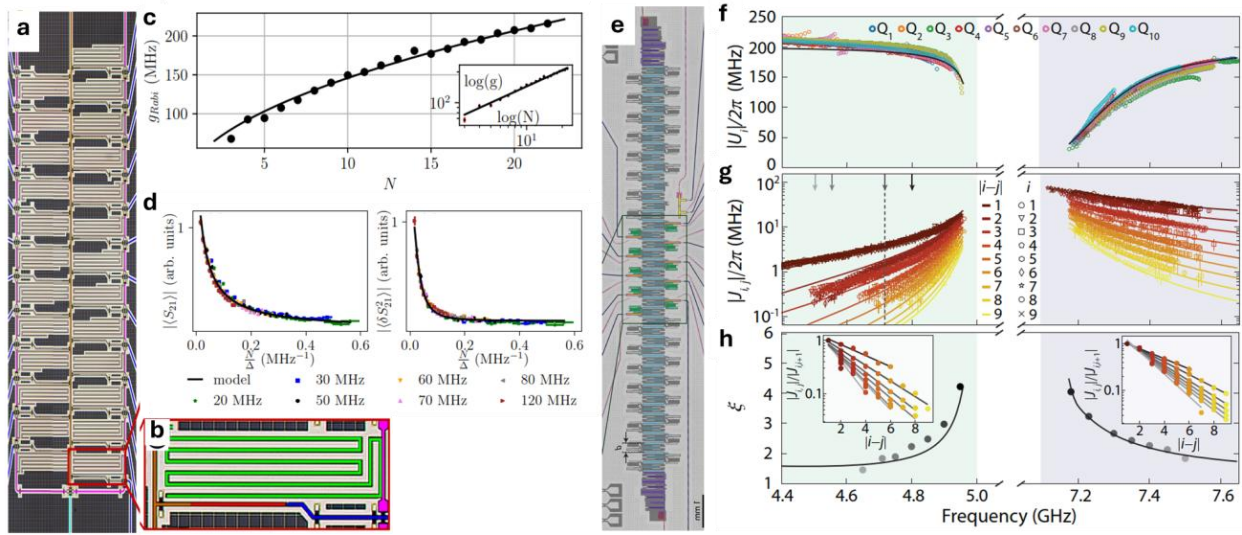

*Figure 1.* (a–d) Device and measurements from a cavity-QED platform with 25 frequency-tunable transmon qubits coupled to a common coplanar resonator. (a) Optical micrograph of the chip layout. (b) Zoom-in showing the individual readout resonators and flux bias lines. (c) Observation of collective bright-state formation through √N scaling of Rabi splitting ($g_{Rabi}$) with the number of resonant qubits N; inset shows power-law fit. (d) Measured average transmission amplitude ⟨|S21|⟩ and mesoscopic fluctuations ⟨|δS21|²⟩ at the cavity frequency as a function of N/Δ for various levels of synthetic frequency disorder Δ, confirming crossover from mesoscopic to thermodynamic behavior. (Panels adapted from Mazhorin et al., Phys. Rev. A 105, 033519 (2022). Copyright 2022 American Physical Society, Creative Commons.)[323]. (e–h) Realization of a superconducting quantum simulator based on a one-dimensional photonic-bandgap metamaterial waveguide with 10 transmon qubits. (e) Device layout with a 42-cell LC resonator array acting as a waveguide with embedded qubits. (f) On-site nonlinearity |Ui| as a function of

qubit frequency, showing strong variation due to hybridization with the band edge. (g) Long-range photon-mediated hopping rates |Jij| measured between qubits at different distances |i–j| and frequencies, illustrating exponential decay with distance. (h) Extracted localization lengths ξ of photonic bound states within the lower and upper bandgaps; insets show decay of normalized couplings |Jij/Ji,i+1| with |i–j| on a log scale. Panels adapted from Zhang et al., Science 379, 278–283 (2023). Copyright 2023 American Association for the Advancement of Science, Creative Commons)[42].

A primary challenge in scaling any quantum technology is overcoming the inherent disorder from fabrication imperfections. A quantum metamaterial platform provides a unique opportunity to study these effects in a controlled manner. Mazhorin et al.[323] demonstrated such a system using a processor with 25 individually frequency-tunable transmon qubits all capacitively coupled to a common coplanar resonator (Fig. 1a, b). This architecture is a direct realization of the Tavis-Cummings model, which describes the collective interaction of many two-level systems with a single light mode. As a foundational benchmark, the experiment first confirmed the coherent nature of the collective coupling by tuning up to N=23 qubits into resonance with the cavity and observing the characteristic $g\sqrt{N}$ scaling of the collective vacuum Rabi splitting, as shown in Fig. 1c.

The primary goal of the work, however, was to use this platform as an analog simulator for a disordered quantum metamaterial. By applying a controlled, statistical spread Δ to the qubit frequencies, the researchers studied the system's response to an external microwave probe. As shown in Fig. 1d, they measured how the average microwave transmission and its variance change as a function of the ratio $N/\Delta$. These results demonstrated a clear transition from a mesoscopic regime, where the system response is dominated by large fluctuations between different disorder realizations, to a self-averaging thermodynamic limit where these fluctuations are suppressed. This work provides a powerful experimental testbed for understanding the robustness of collective quantum phenomena against realistic imperfections and for exploring the fundamental physics of disordered many-body systems.

While coupling qubits to a single mode provides all-to-all connectivity, engineering more structured Hamiltonians requires finer control over the interaction graph. An alternative metamaterial architecture achieves this by coupling qubits to a multi-mode quantum bus with an engineered dispersion relation. Zhang et al.[42] realized such a platform by embedding a one-dimensional array of 10 transmon qubits within a metamaterial waveguide composed of 42 coupled lumped-element resonators (Fig. 1e). This waveguide features a photonic bandgap, a range of frequencies where photon propagation is forbidden. When the qubits are tuned into this bandgap, their excitations dress the surrounding waveguide modes to form localized "qubit-photon bound states". The interactions in this system are mediated by the spatial overlap of the decaying photonic tails of these bound states, realizing an extended Bose-Hubbard model. A key feature of this platform is the exquisite control over the many-body Hamiltonian. As shown in Fig. 1f, the on-site interaction strength, $U_i$, which corresponds to the qubit's anharmonicity, is strongly tunable by changing the qubit's frequency relative to the band edge. Most importantly, the inter-qubit hopping strength, $|J_{i,j}|$, decays exponentially with distance, and its characteristic decay length, $\xi$, can be programmed by adjusting the frequency detuning (Fig. 1g, h). By tuning the qubits closer to the band edge, the localization length of the bound states increases, thereby increasing the interaction range. This programmability allowed the researchers to use the platform to simulate the many-body quench dynamics of the extended Bose-Hubbard model and observe the crossover from an integrable (nearest-neighbor) to a chaotic (long-range) regime.

## 6. Harnessing Topology: Protected States in Metamaterial Waveguides

A paradigm-shifting application of metamaterials in quantum circuits is the ability to move beyond simple electromagnetic engineering and physically instantiate models from condensed matter physics, complete with their non-trivial topological properties[49]. This powerful synergy allows for the exploration of topological phases of matter, which are characterized by global properties that are immune to local perturbations. The central promise of this approach, often termed "topological photonics" when applied to light, is the creation of topologically protected quantum states[324–328]. These states are intrinsically robust against the local sources of noise and disorder – such as fabrication imperfections and stray fields – that represent a primary obstacle in building scalable and reliable quantum computers. By meticulously engineering the band structure of a photonic bath, the topology of the metamaterial can be imprinted onto the quantum emitters coupled to it, leading to novel, protected functionalities for storing and transmitting quantum information. This section delves into this exciting frontier, using the archetypal Su-Schrieffer-Heeger (SSH) model as a case study to illustrate how the abstract concepts of topology can be brought to life in superconducting circuits, yielding tangible experimental signatures and opening new pathways for quantum information processing.

### 6.1 The Su-Schrieffer-Heeger (SSH) Model in Circuit QED

The Su-Schrieffer-Heeger (SSH) model is the canonical one-dimensional model of a topological insulator, originally developed to describe the behavior of electrons in the long-chain polymer polyacetylene[329]. Its elegant simplicity and rich physical consequences have made it a cornerstone for understanding topological phenomena in a vast array of physical systems, from photonics and acoustics to mechanical metamaterials and, most recently, circuit QED[330–336]. The model describes spinless particles hopping on a one-dimensional lattice that is "dimerized," meaning it is composed of a repeating unit cell containing two distinct sites, labeled A and B. The hopping amplitude between sites within the same unit cell (intracell) is different from the hopping amplitude between adjacent sites in neighboring unit cells (intercell)[49]. In the tight-binding approximation, the Hamiltonian for the SSH model is given by: $\hat{H}_{SSH} = \sum_n ( v\hat{a}_{n,B}^\dagger \hat{a}_{n,A} + w\hat{a}_{n+1,A}^\dagger \hat{a}_{n,B} + \text{h.c.})$. Here, $\hat{a}_{n,A}^\dagger$ ($\hat{a}_{n,A}$) is the creation (annihilation) operator for a particle on site A of the $n$-th unit cell, and similarly for site B. The parameter $v$ represents the intracell hopping amplitude, while $w$ represents the intercell hopping amplitude. The crucial physics of the SSH model arises from the competition between these two hopping strengths. The model possesses a fundamental chiral (or sublattice) symmetry, which dictates that the Hamiltonian only contains terms that couple A-sites to B-sites, and never A-to-A or B-to-B. This symmetry is responsible for the emergence of two distinct topological phases.

The topological character of the chain is determined by the choice of dimerization, i.e., the relative magnitude of $v$ and $w$. When the intracell coupling is stronger ($|v| > |w|$), the system is in a *topologically trivial phase*. In this configuration, the chain can be thought of as a collection of strongly coupled dimers that are weakly connected to each other. When the intercell coupling dominates ($|w| > |v|$), the system enters a *non-trivial topological phase*. Here, the strong bonds connect different unit cells, fundamentally changing the global structure of the chain. This distinction is not merely a matter of convention; it is a profound physical difference that is mathematically captured by a topological invariant. For one-dimensional systems with inversion symmetry, this invariant is the *Zak phase*. The Zak phase is a specific type of Berry phase, which quantifies the geometric phase acquired by a particle's Bloch wavefunction as it traverses the entire Brillouin zone[337]. For a system with inversion symmetry, the Zak phase is quantized to one of two values: 0 for the trivial phase or $\pi$ for the topological phase. The phase transition between the two regimes occurs precisely when $|v| = |w|$, at which point the energy gap between the two bands of the system closes.

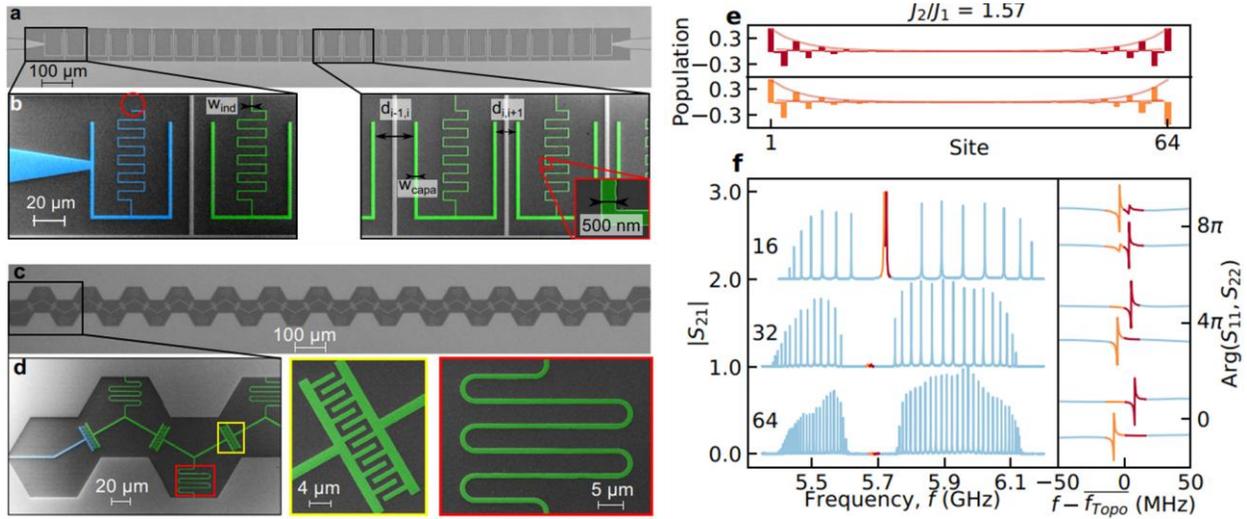

*Figure 2.* Topological protection and edge-state localization in superconducting quantum metamaterials. (a–b) SEM images of a superconducting transmission line metamaterial implementing a Su-Schrieffer–Heeger (SSH) model with alternating capacitive couplings. The unit cells contain coplanar waveguide resonators with engineered coupling distances ($d_{i-1,i}$, $d_{i,i+1}$) to set intra- and inter-cell coupling strengths. (c–d) Device layout of a modified SSH lattice with corner and defect resonators for enhanced control; zoom-ins show sub-micron feature sizes of capacitive and inductive elements. (e) Experimental measurement of photon population along the lattice, revealing exponential localization of edge modes at site 1 and site 64 for $J_2/J_1 = 1.57$. (f) Transmission spectra $|S_{21}|$ for increasing chain lengths (16, 32, 64), showing emergence of a topological mid-gap mode; right: phase winding analysis of scattering matrix elements confirms quantized topological signature. Adapted from Jouanny et al., Nat. Commun. 16, 3396 (2025). Copyright 2025 Nature Portfolio, Creative Commons.[140]

The true power of circuit QED is its ability to realize such abstract Hamiltonians with high fidelity and control[334]. The photonic analog of the SSH model is constructed using a metamaterial waveguide composed of a one-dimensional array of superconducting LC resonators[334]. Each resonator acts as a photonic site (A or B). The hopping of microwave photons between adjacent resonators is mediated by engineered capacitive and/or inductive couplings. The crucial dimerization is achieved by physically alternating these coupling strengths along the chain – for example, by using two different values for the coupling capacitors, $C_v$ and $C_w$, to realize the hopping amplitudes $v$ and $w$. To probe the rich physics of this topological waveguide, superconducting qubits, such as transmons, are capacitively coupled to individual resonator sites. These qubits act as highly sensitive, in-situ quantum emitters, allowing for local excitation and spectroscopic measurement of the waveguide's properties.

For example, the work in Fig.2 illustrates how superconducting circuit architectures can emulate topological models and probe protected quantum states. The metamaterial is fabricated as a 1D chain of microwave resonators with alternating strong and weak capacitive couplings to realize an SSH-type tight-binding model. The SEM panels show meticulous engineering of circuit parameters to control hopping amplitudes. The localization of photonic wavefunctions at the edges, quantified in Fig.2e, demonstrates the presence of symmetry-protected zero-energy edge states, a hallmark of topological phases. The transmission spectra in Fig.2f further reveal the resilience of these states to increasing system size, and the associated topological invariants are extracted from the complex S-matrix argument, confirming a nontrivial winding number. These results highlight the feasibility of exploring topological quantum optics using scalable and lithographically programmable superconducting platforms.

### 6.2 Experimental Signatures: Probing Topology with Light

The abstract mathematical distinction between topological phases manifests in concrete, measurable physical phenomena. The most celebrated hallmark of a non-trivial topological phase in a finite-sized SSH chain is the emergence of *topologically protected edge states*[338–340]. These are special solutions to the Hamiltonian that exist at zero energy (i.e., in the middle of the bulk bandgap) and whose wavefunctions are exponentially localized at the two ends of the chain. Their existence is guaranteed by the bulk-boundary correspondence principle, which states that a non-trivial topological invariant in the bulk of a material necessitates the existence of protected states at its boundary[49]. These edge states are remarkably robust; their existence and zero-energy property are protected by the chiral symmetry of the lattice and are immune to local perturbations and disorder that do not break this fundamental symmetry. In the circuit QED implementation, these topological features are probed by coupling a qubit to the metamaterial waveguide. When the qubit's transition frequency is tuned to lie within the photonic bandgap of the waveguide, it cannot radiate into propagating modes. Instead, the qubit and the waveguide photons form a hybrid, localized excitation known as a *qubit-photon bound state*[41,341]. The photonic component of this bound state can be visualized as a "cloud" of virtual photons whose spatial profile is dictated by the waveguide's eigenmodes.

This leads to a remarkable and unambiguous experimental signature: *directional emission*[334,339]. In the trivial phase ($|v| > |w|$), the photon cloud of the bound state is symmetric and extends equally in both directions away from the qubit. However, in the topological phase ($|w| > |v|$), the photon cloud's wavefunction takes on the character of the underlying topological edge state. This means its spatial profile becomes highly asymmetric and exponentially localized towards one end of the chain. Crucially, the direction depends on whether the qubit is coupled to an A- or B-site. Measurements of emission from external microwave ports confirm this chiral behavior, with strong output on one side and near-complete suppression on the other – direct evidence of the waveguide's nontrivial topology. Beyond the bandgap, the waveguide's topology also shapes photon scattering in the passband. Qubit interactions, mediated by real photons, depend on the phase accumulated during propagation, this phase is topology-dependent. As a result, the transmission spectrum shows distinct interference patterns, including "perfect superradiance" points where coherent exchange vanishes. The number of such points differs by one between trivial and topological phases, offering a clear spectroscopic signature of topological order[334].

## 6.3 Prospects for Topologically Protected Quantum Information Processing

The ultimate ambition of *topological quantum computation* (TQC) is to encode quantum information in the non-local, degenerate ground states of a topologically ordered system[342–346]. In such a scheme, logical operations are performed by braiding *non-Abelian anyons* – exotic quasiparticles whose exchange statistics are non-commutative. Because the information is stored globally, it is intrinsically immune to local errors, offering a tantalizing path toward fault-tolerant quantum computing. While the 1D SSH model realized in circuit QED does not possess the non-Abelian statistics or degenerate ground states needed for full fault-tolerant quantum memory, it serves as a critical and highly successful experimental testbed for the foundational principles of topological physics in an engineered quantum system[334,342]. The ability to design, fabricate, and probe these topological states with single-site resolution provides an invaluable platform for verifying theoretical models and developing the techniques that will be necessary for more complex topological architectures.

The most immediate application of these systems is in the realm of quantum communication. The robust, directional nature of the topological edge states can be harnessed to create a protected quantum channel[334]. A protocol has been experimentally demonstrated where a quantum state is faithfully transferred from a "sender" qubit at one end of the chain to a distant "receiver" qubit at the other end. This process is mediated by the coherent exchange of population between the two topological edge states of the waveguide. The protocol proceeds in three steps: (1) an iSWAP gate transfers the excitation from the sender qubit to the left-localized edge state; (2) the two edge states, which have a small but finite overlap in the center of the chain, coherently exchange the excitation; (3) a final iSWAP gate maps the population from the right-localized edge state to the receiver qubit. This demonstration, achieving a state transfer fidelity of 87%, represents a significant step beyond simple nearest-neighbor coupling, enabling robust, long-range quantum state transfer that is protected from disorder in the intervening chain.

Furthermore, these circuit QED platforms open the door to simulating more complex topological phenomena. The zero-energy edge states of the SSH model are, in many ways, the simplest analogue of *Majorana zero modes (MZMs)*, which are predicted to exist at the ends of 1D topological superconductors[344,347–353]. MZMs are quasiparticles that are their own antiparticles and are predicted to obey non-Abelian braiding statistics, making them a primary candidate for realizing TQC. Although definitive experimental confirmation of MZMs and their braiding properties remains elusive, recent advances in hybrid superconductor–semiconductor nanowire platforms have enabled single-shot interferometric measurements of fermion parity consistent with the presence of MZMs[354]. These results mark substantial progress toward controlled parity readout and measurement-based topological operations, even though the topological nature of the observed states cannot yet be unambiguously distinguished from finely tuned trivial bound states. Recent proposals suggest that classical mechanical metamaterials, which can also be designed to exhibit an SSH-like topological phase, could be used to explore the braiding statistics of classical analogues of MZMs[355–357]. By demonstrating the fundamental building blocks of topological protection in a highly controllable and scalable circuit QED environment, these experiments pave the way for future architectures that may realize more advanced topological codes, protected quantum operations, and ultimately, a fault-tolerant quantum computer.

## 7. Exotic Excitations for Quantum Control and Coherence

Beyond engineering band structures and impedance, the design freedom afforded by metamaterials offers the possibility of creating and controlling exotic electromagnetic excitations that have no counterpart in simple circuits or natural materials[291,358–366]. By designing meta-atoms with specific symmetries and current distributions, it is possible to access novel light-matter interaction pathways that fundamentally alter how a quantum system couples to its environment. This opens up revolutionary strategies for protecting quantum information not by incrementally fighting decoherence, but by designing qubits that are intrinsically immune to dominant noise channels[39,41,42,52,140,334,367–371]. This frontier of research moves beyond passive environmental engineering towards the active creation of new quantum objects with built-in protection. Two of the most promising and conceptually rich frontiers in this area are the use of non-radiating anapole modes to create "quiet" qubits and the development of entirely new qubit paradigms based on the topological stability of artificial magnetic skyrmions.

### 7.1 Toroidal Dipoles and Anapole Modes: Suppressing Radiative Decay for "Quiet" Qubits

The interaction of matter with light, and consequently the process of spontaneous emission, is typically described by a multipole expansion of the charge-current distribution of the emitter. In most systems, this expansion is overwhelmingly dominated by the lowest-order terms: the electric dipole moment (**P**) and the magnetic dipole moment (**M**). These moments are the primary sources of radiation, and therefore, radiative decay into the electromagnetic vacuum (the Purcell effect) is a fundamental and often dominant decoherence channel for any quantum emitter, including a superconducting qubit[307,367,372–375]. The core strategy of many coherence-enhancing techniques is to suppress this interaction, for example by placing the qubit in a cavity or a photonic bandgap material to reduce the density of available electromagnetic modes.

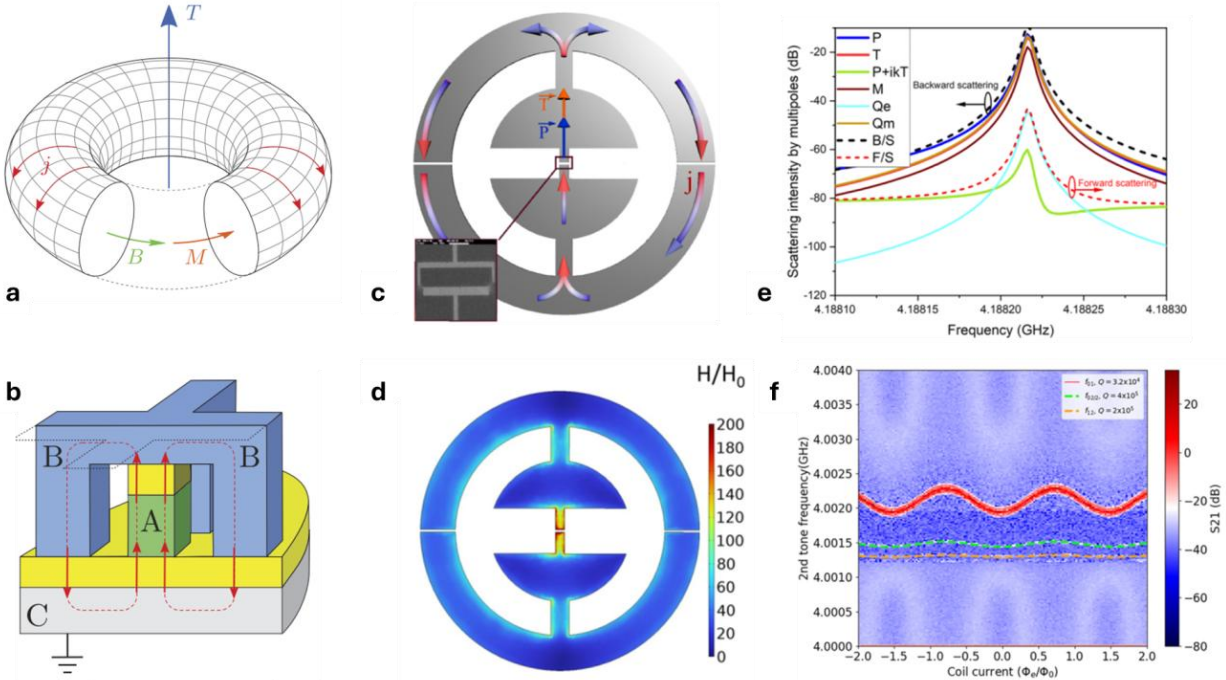

*Figure 3*. Toroidal superconducting qubits: geometric decoupling and multipolar scattering signatures. (a) Conceptual sketch of a toroidal current distribution generating a toroidal dipole moment **T**, magnetization **M**, and no net electric or magnetic dipole radiation. (b) 3D schematic of a "closed" toroidal superconducting qubit based on stacked Josephson junction layers (A, B) and substrate (C). Adapted from Zagoskin et al., Sci. Rep. 5, 16934 (2015). Copyright 2015 Nature Publishing Group, Creative Commons[376]. (c) Realization of a toroidal resonator with azimuthal currents; inset: SEM of the nanofabricated structure. (d) Simulated magnetic field amplitude $H/H_0$ showing strong field localization in the center of the toroid. (e) Scattering decomposition into multipoles showing dominant toroidal dipole and magnetic quadrupole contributions, with suppressed forward scattering due to anapole-like interference. (f) Two-tone spectroscopy of a tunable toroidal cavity-qubit system, showing multiple high-Q modes and strong flux dispersion near $f_{12}$. Adapted from Stenishchev et al., Phys. Rev. B 110, 035157 (2024). Copyright 2024 American Physical Society, Creative Commons[377].

However, a more radical approach exists: designing an emitter whose fundamental coupling to the electromagnetic field is not dipolar. The standard multipole expansion is, in fact, incomplete. A third, independent family of electromagnetic multipoles exists: the *toroidal multipoles*, which were first considered in the context of parity violation in nuclear physics[378]. The lowest-order toroidal dipole arises from poloidal current loops confined within a toroidal geometry, producing no external magnetic field but a nonzero vector potential[379–381]. In scattering terms, it corresponds to a resonant eigenmode with unique symmetry – odd under both parity and time reversal – whose far-field radiation can cancel with electric dipole contributions, enabling dark or weakly radiating modes that are spectrally distinguishable from conventional multipoles[294]. The true power of the toroidal dipole emerges when it is considered in conjunction with a conventional electric dipole. A remarkable phenomenon, known as an *anapole mode*, occurs when an electric dipole moment, **P**, and a toroidal dipole moment, **T**, are co-located and oscillate with a specific, out-of-phase relationship[292,293,382,383]. Although each dipole radiates individually, their far-field radiation patterns are sufficiently similar that they can destructively interfere and cancel each other out completely. This perfect cancellation, which renders the source non-radiating and effectively invisible to the far field, is achieved when the moments satisfy the condition: $\mathbf{P} - ik\mathbf{T} = 0$, where $k$ is the wave number of the radiation[383]. A system supporting such a mode does not radiate energy, effectively closing the primary channel for Purcell decay[384]. It is important to note that while an anapole does not radiate an electromagnetic field, it can still possess a non-trivial vector potential, which can lead to observable phenomena such as the Aharonov-Bohm effect.

This principle can be directly harnessed for quantum technologies by designing metamaterial resonators whose fundamental resonance is an anapole mode. While realizing the ideal toroidal geometry is challenging, planar metamaterial structures can be engineered to exhibit anapole behavior[385,386]. A prominent example consists of a metamolecule formed by two symmetric split-ring resonators (SRRs). When an incident plane wave excites the structure, it induces circulating currents in the SRR loops. These currents generate circulating magnetic moments that, in turn, produce a strong toroidal dipole moment oscillating along the metamolecule's axis. Simultaneously, a central gap in the structure allows for the excitation of a conventional electric dipole moment. By carefully tuning the geometry, the amplitudes and phases of these two moments can be balanced to satisfy the anapole condition, leading to a sharp suppression of far-field scattering and a dramatic increase in the resonator's quality factor[291]. Such non-radiating anapole modes have been experimentally demonstrated in dielectric nanoparticles and all-dielectric metamaterials, confirming their potential for creating ultra-high-Q resonators.

The concept of anapole modes inspires a novel design paradigm for qubits themselves: the *toroidal qubit*[376,387]. This would be a superconducting circuit, such as a flux qubit, engineered with specific symmetries to ensure that its coupling to the environmental electromagnetic field is dominated by its toroidal moment, while its electric and magnetic dipole moments are suppressed by design[387]. Such a qubit would be naturally "quiet," as it would be intrinsically decoupled from the most common environmental noise channels that couple to electric and magnetic dipoles. This offers a path to enhanced coherence times, not by shielding the qubit, but by making it inherently stealthy. Proposed designs for such qubits often involve multi-junction loops that support circulating Josephson currents, creating the necessary poloidal current flow to generate a net toroidal moment. The interaction Hamiltonian for a toroidal qubit would be fundamentally different from the standard capacitive or inductive coupling Hamiltonians, coupling to gradients of the external field (e.g., $\mathbf{T} \cdot (\nabla \times \mathbf{H})$) and offering a new, protected way to manipulate quantum states. The successful demonstration of anharmonicity in a superconducting anapole meta-atom with an embedded Josephson junction, yielding an anharmonicity of 540 kHz, underscores the promise of this approach for building robust quantum computing elements[377].

Figure 3 highlights the emergence and functionality of toroidal superconducting qubits as "quiet" artificial atoms that couple predominantly through toroidal moments. Unlike conventional dipolar qubits, the toroidal architecture ensures vanishing electric and magnetic dipole moments at leading order, leading to anapole behavior that suppresses radiation losses and mitigates coupling to low-frequency environmental noise. Figs.3 (a) and (b) illustrate both the theoretical topology and practical implementation of closed-loop current distributions engineered via stacked Josephson junctions[376]. In Figs.3(c–d), experimental structures and simulations confirm that magnetic energy is strongly confined within the central toroidal region, enhancing non-radiative modes. The scattering analysis in Fig.3e demonstrates that the toroidal moment dominates over electric and magnetic dipoles, enabling destructive interference in the forward channel, a hallmark of anapole physics. Finally, spectroscopy data Fig.3f shows that such cavity-embedded toroidal systems support well-resolved resonances with high quality factors and flux sensitivity, indicating their potential utility in tunable quantum circuits[377]. Collectively, these results position toroidal qubits as a promising route toward decoherence-resistant quantum elements with intrinsically weak environmental coupling.

## 7.2 Artificial Skyrmions: A New Qubit Paradigm from Topological Magnetism

While toroidal qubits seek to achieve coherence by minimizing interactions with the environment, an entirely different paradigm aims to achieve robustness through intrinsic topological stability. This approach draws inspiration from the field of magnetism and leverages a fascinating class of quasiparticles known as *magnetic skyrmions*[362,363,363,388–390]. Skyrmions are nanoscale, particle-like swirls or textures in the magnetization of a material. Their defining feature is their topology: the spin configuration of a skyrmion has a non-trivial winding number, typically an integer value $Q$, which quantifies how many times the spin vectors wrap around a unit sphere as one traverses the texture. A simple ferromagnetic state has $Q = 0$, while a skyrmion has $|Q| \geq 1$. This non-trivial topology means that a skyrmion cannot be continuously "unwound" into a uniform ferromagnetic state without overcoming a significant energy barrier. This *topological protection* makes skyrmions exceptionally stable and resilient to local perturbations, such as thermal fluctuations or material defects – a highly desirable feature for any form of information storage. Skyrmions were first observed in bulk materials lacking inversion symmetry, where the Dzyaloshinskii-Moriya interaction (DMI) stabilizes their chiral structure[391–393]. However, it is also possible to create *artificial* skyrmion crystals in metamaterials that do not possess intrinsic DMI, for instance, by nanopatterning arrays of magnetic vortices onto a perpendicularly magnetized film[394,395]. This ability to engineer skyrmionic systems opens the door to their use in quantum technologies.

This stability has inspired a new paradigm for quantum information: the *skyrmion qubit*[396–398]. Instead of encoding information in the charge or flux of a superconducting circuit, a skyrmion qubit encodes it in a collective, quantum degree of freedom of the spin texture itself. The leading proposal uses the skyrmion's *helicity*, $\phi_0$, as the basis for the qubit[398]. The helicity describes the sense of rotation of the in-plane spins within the texture – for example, whether they rotate clockwise or counter-clockwise from the core to the edge. In a bulk material, the helicity is often fixed by the DMI. However, in frustrated magnets or geometrically confined nanostructures, the helicity can become a dynamic degree of freedom. When a nanoscale skyrmion is confined, its helicity, which is a continuous variable classically, becomes quantized due to quantum tunneling between different helicity configurations. This quantization gives rise to a discrete energy spectrum that can be used to define a qubit. This platform offers a remarkable degree of control. The qubit's transition frequency, $\omega_q$, and its anharmonicity can be tuned *in situ* by applying external electric and magnetic fields, providing a rich operational regime. Single-qubit gates can be performed by applying microwave-frequency electric fields or spin currents, which couple to the skyrmion's helicity and drive rotations on the Bloch sphere. The driven Hamiltonian in the rotating frame takes the standard form for qubit control[399]: $\hat{H}_{rot} = \frac{\Delta\omega}{2}\hat{\sigma}_z + \frac{\Omega(t)}{2}[\cos(\varphi)\hat{\sigma}_x + \sin(\varphi)\hat{\sigma}_y]$, where $\Delta\omega$ is the detuning from the qubit frequency and $\Omega(t)$ is the Rabi frequency controlled by the microwave drive.

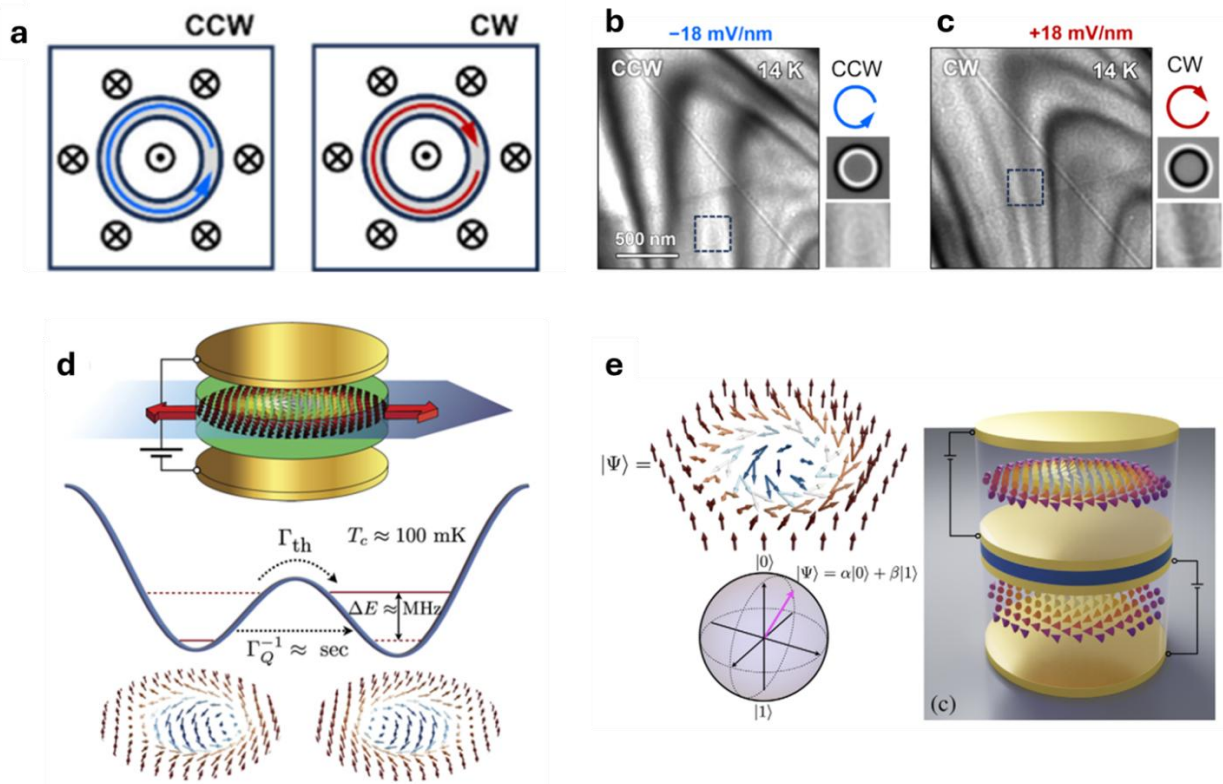

**Figure 4.** From classical control to quantum logic with skyrmion helicity. (a) Calculated in-plane spin circulation for counter-clockwise (CCW, blue) and clockwise (CW, red) Bloch-type skyrmionic bubbles in centrosymmetric $Cr_2Ge_2Te_6$, highlighting the two helicity eigenstates. (b,c) Cryogenic Lorentz-TEM images recorded at 14 K after field-cooling in −18 mV nm⁻¹ (b) and +18 mV nm⁻¹ (c) out-of-plane electric fields. Reversing the field deterministically selects CCW or CW helicity while leaving polarity unchanged, evidencing purely electric control of the order-parameter. Adapted from Han et al., Nano Lett. 25, 5174 (2025). Copyright 2025 American Chemical Society[400]. (d) Low-temperature energy landscape for a single skyrmion helicity degree of freedom in a frustrated magnet: anisotropy and electric gating create a double-well potential with barrier height giving an inverse macroscopic quantum-tunnelling time $\Gamma_Q^{-1} \approx 1$ s below ≈ 100 mK and level splitting $\Delta E \approx$ MHz. Adapted from Psaroudaki et al., Phys. Rev. B 106, 104422 (2022). Copyright 2022 APS, Creative Commons[398]. (e) Qubit blueprint in which the two helicity states $|0\rangle$ = CCW and $|1\rangle$ = CW form a Bloch-sphere basis; microwave magnetic-field gradients and gate electrodes in a vertical heterostructure provide manipulation and dispersive read-out pathways. Adapted from Psaroudaki et al., Phys. Rev. Lett. 127, 067201 (2021). Copyright 2021 APS, Creative Commons[397].

For scalability, two-qubit gates are essential. A straightforward scheme involves the interlayer exchange interaction in a magnetic bilayer, where two skyrmion-hosting layers are separated by a thin nonmagnetic spacer. This mediates an interaction between the helicities of the two skyrmions, described by a coupling Hamiltonian of the form[399]: $H_{int} = -J_{ex}\cos(\phi_1 - \phi_2)$. This interaction can be used to implement controlled-phase gates. An alternative and highly promising route for long-range coupling involves mediating the interaction with a third quantum system, such as a nanomechanical resonator. In such a hybrid system, the skyrmion qubit couples magnetically to the phonons of the resonator, which can then act as a quantum bus to connect distant skyrmion qubits[401].

The theoretical foundation for this scheme, developed primarily by Psaroudaki and Panagopoulos[397,398], treats helicity as a quantum variable subject to a potential landscape $\varphi_0$. By introducing in-plane magnetic anisotropy achievable through mechanisms such as strain or engineered defects one can break the continuous rotational symmetry and generate a double-well potential favoring discrete helicity states. An external out-of-plane electric field then allows for control over the potential barrier height and asymmetry, thereby tuning the energy splitting between the logical states. Within this framework, the skyrmion helicity system is predicted to exhibit macroscopic quantum phenomena at low temperatures (below ~100 mK). Theoretical models indicate that for skyrmions with radii on the order of 5 nm, quantum tunneling between the two helicity states leads to an energy splitting ΔE in the MHz range, while the inverse of the tunneling rate can be on the order of seconds. This macroscopic quantum tunneling (MQT) sets the stage for coherent oscillations between helicity states, enabling quantum operations[398], Fig. 4d. Control protocols have also been proposed: single-qubit gates via microwave magnetic field gradients, and tunable two-qubit interactions through interlayer exchange coupling in vertical magnetic heterostructures. Projected coherence times lie in the microsecond regime, primarily limited by Gilbert damping. Achieving the scalable device architecture[397] outlined in Fig. 4e will require both improved material quality and experimental confirmation of coherent tunneling dynamics. Thus, while the skyrmion qubit has evolved from theoretical speculation to a prototype with demonstrated state control, its practical utility as a quantum logic element remains contingent on bridging the persistent gap between theoretical possibility and experimental control.

Despite its conceptual elegance, the skyrmion qubit's feasibility depends critically on the ability to deterministically manipulate helicity states. A recent advance by Han et al. has achieved precisely that[400]. Investigating the van der Waals magnet $Cr_2Ge_2Te_6$, centrosymmetric and lacking intrinsic DMI, they observed that skyrmion helicities are normally random. Using in situ Lorentz transmission electron microscopy (LTEM), they demonstrated that applying an out-of-plane electric field during field-cooling breaks inversion symmetry and induces a finite DMI. Importantly, the sign of the electric field determines the sign of the induced DMI, thereby selecting the helicity of the skyrmions that form. Reversing the electric field switches the population between CCW and CW helicity states. Figure 4 a–c confirm that moderate gate voltages during field-cooling of $Cr_2Ge_2Te_6$ reliably determine the sign of DMI, enabling deterministic selection between CW and CCW skyrmions. These electrically programmed states remain stable at cryogenic temperatures after field removal, offering a robust method to write the logical basis states of the proposed qubit.

## 8. Advanced Qubit Architectures with Metamaterial Resonators

As the field of quantum computing matures, the focus is shifting from improving individual components to designing robust, fault-tolerant architectures capable of solving classically intractable problems. The state-of-the-art error rates for physical qubits, while impressive, remain many orders of magnitude higher than what is required for large-scale algorithms[12,402,403]. This gap necessitates the use of quantum error correction (QEC), where information is encoded redundantly to protect it from noise[12,404–408]. Metamaterial resonators play an important role in this transition, evolving from passive circuit elements into active computational resources that enable novel and highly efficient QEC schemes. This is most evident in the development of bosonic quantum codes, which leverage the unique properties of engineered electromagnetic environments to encode and protect quantum information in a hardware-efficient manner, and in inspiring entirely new qubit paradigms based on topological principles[405,407].

**Table 2**: State-of-the-Art Error Rates for Physical Qubits.

| Platform | Single-Qubit Gate Error Rate | Two-Qubit Gate Error Rate | Ref. |
| --- | --- | --- | --- |
| **Trapped-Ion (Ca⁺)** | < 1 × 10⁻⁷ (0.00001 %) | – | [402] |
| **Superconducting Transmon** | (7.42 ± 0.04) × 10⁻⁵ (~0.0074 %) | – | [409] |
| **Superconducting Fluxonium** | 0.002 % | 0.08 % | [410] |
| **Diamond NV Center Spin Qubit** | < 0.1 % | < 0.1 % | [411] |
| **Neutral-Atom Rydberg Qubit** | < 0.03 % | 0.5 % | [412] |
| **Silicon Spin Qubit (Si/SiGe)** | < 1 % | < 1 % | [413] |

Conventional approaches to QEC, such as the well-studied surface code, rely on massive redundancy, using many physical two-level qubits to encode a single, more robust logical qubit. For instance, a recent experimental demonstration required approximately 50 physical qubits to create a single logical qubit that reached the "break-even" point, where the lifetime of the logical qubit exceeds that of its best physical constituent[414]. While powerful, the immense hardware overhead of this approach presents a daunting scaling challenge for building a useful quantum computer.

An alternative and highly promising paradigm, known as *bosonic quantum error correction*, offers a path to fault tolerance with significantly greater hardware efficiency. Instead of encoding information across many two-level systems, bosonic QEC encodes a logical qubit into the vast, infinite-dimensional Hilbert space of a single physical system, such as the quantized electromagnetic modes of a high-Q superconducting resonator[11,404,407,415]. This approach leverages the fact that a single harmonic oscillator provides a naturally expansive state space in which to build redundancy, potentially reducing the physical footprint and complexity of a logical qubit dramatically. The fundamental principle of bosonic QEC is to define logical qubit states, $|\bar{0}\rangle_L$ and $|\bar{1}\rangle_L$, as specific, highly non-classical superpositions of the oscillator's Fock states (number states, $|n\rangle$). These superpositions are carefully designed to have specific symmetries that make them robust against the dominant error channels of the physical system. For a high-Q resonator, the primary error is photon loss, where a single photon leaks out into the environment. Different bosonic codes are tailored to combat different types of errors. For example, "*cat codes*" are designed to be robust against single-photon loss[404,405,407,408], while *Gottesman-Kitaev-Preskill (GKP)* codes are designed to correct for small, continuous shifts in the oscillator's position and momentum quadratures[416–418].

The primary challenge in implementing bosonic codes is that preparing, manipulating, and measuring these complex, non-Gaussian quantum states requires strong nonlinear interactions. A simple linear resonator cannot, on its own, create the necessary superpositions or perform the required logic. This is precisely where the synergy between a high-Q linear resonator (the "boson") and a nonlinear element (the "ancilla," typically a transmon qubit) becomes crucial. The transmon ancilla, strongly coupled to the resonator, provides the necessary nonlinearity to control the bosonic state, prepare the encoded logical states, and perform non-destructive measurements of error syndromes. Metamaterial resonators can enhance this architecture by providing engineered mode structures and impedances that optimize the interaction between the ancilla and the bosonic mode, making the entire process more efficient and controllable[11,419].

## 9. Outlook: Scalability, Integration, and Novel Frontiers

The integration of metamaterial principles into superconducting quantum circuits has opened a vast design space, providing powerful new tools to address the most pressing challenges facing the development of large-scale, fault-tolerant quantum computers. The journey from demonstrating single-qubit control to building processors with hundreds or thousands of qubits is not merely a matter of repetition; it is a complex, multi-faceted engineering and physics problem that demands new architectural paradigms. The future of the field lies in leveraging the unique capabilities of metamaterials to solve the interconnected problems of scalability, connectivity, and coherence, while simultaneously pushing the boundaries of physics with novel hybrid systems and design methodologies.

### 9.1 Metamaterials as a Solution to the Wiring and Connectivity Bottleneck

A critical and formidable obstacle to scaling up superconducting quantum processors is the sheer physical complexity of control and readout wiring[33,420,421]. In conventional architectures, each individual qubit requires several dedicated microwave lines for control (e.g., XY-drives for rotations, Z-drives for frequency tuning) and for state readout via a coupled resonator. These lines must be routed from room-temperature electronics down through the various stages of a dilution refrigerator to the quantum processor operating at millikelvin temperatures. This creates an unsustainable "wiring bottleneck" that limits the number of qubits on a single chip and imposes a significant heat load on the cryostat, straining its cooling capacity. The scaling properties of this approach are overwhelmingly poor, far exceeding the constraints described by Rent's Rule for classical integrated circuits, where the number of external connections grows sub-linearly with the number of logic gates[422]. In current quantum processors, the requirement of multiple connections per qubit represents a fundamental barrier to reaching the millions of qubits necessary for fault-tolerant computation.

Metamaterial resonators offer an elegant and potentially paradigm-shifting solution to this scaling problem through frequency-based addressing and engineered connectivity[52,423]. The core idea is to move away from a model where each qubit has its own dedicated control infrastructure to one where a single, shared, multi-mode resource – the metamaterial – can control an entire array of qubits[423]. This is achieved by coupling an array of fixed-frequency qubits to the different unit cells of a multi-mode metamaterial resonator. Because of the engineered dispersion of the metamaterial, it can support a dense spectrum of orthogonal standing-wave modes within a compact physical footprint. Each qubit, by virtue of its position in the lattice, couples with varying strength to these different modes, depending on whether it is located at a voltage antinode or node of a particular standing wave. This architecture enables a powerful control scheme. Instead of requiring a separate drive line for each qubit, a single, high-bandwidth arbitrary waveform generator can be coupled to the entire metamaterial lattice. By synthesizing a drive signal composed of a specific combination of the metamaterial's eigenfrequencies, one can selectively excite a superposition of modes. Through careful choice of the mode amplitudes and phases, these modes can be made to interfere constructively at the location of a single target qubit and destructively everywhere else. This creates a large, localized microwave field at the target qubit, inducing a significant AC Stark shift in its transition frequency. The magnitude of this shift, $\Delta\omega_{Stark}$, for a drive with amplitude $\Omega$ detuned by $\delta$ from the qubit transition, is given by: $\Delta\omega_{Stark} \propto \frac{|\Omega|^2}{\delta}$. This frequency shift can be used to effectively turn a qubit "on" or "off" for gate operations, bringing it into or out of resonance with other qubits or control fields, without requiring any local flux-tuning lines. This approach dramatically reduces the I/O requirements for a quantum processor, mitigating the wiring bottleneck and associated heat load, and thus represents a viable path toward controlling large, frequency-addressable qubit arrays.

Furthermore, metamaterials solve the critical problem of limited connectivity[23,42,52]. Standard architectures often feature only nearest-neighbor coupling between qubits, which is inefficient for implementing many important quantum algorithms and error-correction codes that require long-range interactions[424]. Executing a gate between two distant qubits in such an architecture requires a series of SWAP operations, which increases circuit depth and introduces additional errors. Metamaterial resonators, particularly those with engineered dispersion like left-handed ring resonators, can act as a quantum bus, mediating tunable, long-range interactions between any pair of qubits coupled to the bus[52,424]. The dense mode spectrum of these resonators provides multiple channels for virtual photon exchange between distant qubits. The strength and even the sign of the resulting interaction (e.g., the parasitic ZZ interaction) can be tuned by small adjustments to the qubit frequencies relative to the dense forest of resonator modes. This capability enables programmable, all-to-all connectivity within a qubit module, a significant advantage for algorithmic efficiency and the implementation of advanced, hardware-efficient QEC codes that benefit from higher connectivity.

### 9.2 Challenges and Opportunities in 3D Quantum Integration

To move beyond the limitations of single-chip planar layouts and accommodate the millions of qubits required for fault tolerance, the field is increasingly turning to 3D integration. This approach, inspired by modern semiconductor manufacturing, involves stacking and bonding multiple chip layers together. This modularity allows for a separation of functionalities: for instance, placing the delicate, high-coherence qubits on one chip (the "qubit chip") and the dense, potentially noisy classical control and readout electronics on another (the "interposer" or "control chip"). This separation can improve fabrication yield, as different layers can be optimized and tested independently, and can reduce on-chip crosstalk by physically distancing the qubits from sources of noise.

However, 3D integration introduces its own set of formidable challenges that must be overcome to maintain, let alone improve, overall system performance. First, maintaining high qubit coherence across the stack is paramount[275,425,426]. The dielectric materials used for bonding the chips together and as interlayers can be significant sources of microwave loss, directly degrading qubit lifetimes. Rigorous simulations have shown that introducing a lossy dielectric interlayer decreases the qubit quality factor and, consequently, its relaxation time. The loss tangent, $\tan\delta$, of these materials is a critical parameter, and developing ultra-low-loss bonding agents and dielectrics compatible with cryogenic temperatures is an active area of materials research. Second, high-precision alignment between the stacked chips is required to ensure proper coupling between components on different layers, such as between a qubit on one chip and its readout line on another. This alignment often needs to be on the scale of nanometers, a significant engineering challenge for wafer-scale processes[427–429]. Furthermore, the vertical interconnects themselves, such as indium bumps used in flip-chip bonding, must be designed to be superconducting and low-loss to avoid introducing new decoherence channels. Third, while 3D integration can help isolate components, it also introduces new pathways for unwanted electromagnetic crosstalk between layers, which must be carefully modeled and mitigated through shielding and grounding strategies[279,425]. Additionally, managing the thermal budget of a dense 3D stack is complex. The heat generated by control electronics on one layer can propagate to the qubit layer, increasing the effective temperature and introducing thermal noise that dephases the qubits.

Despite these hurdles, the potential benefits are driving significant research. Metamaterial concepts can be applied to address these 3D integration challenges. For example, engineered structures can serve as vertical interconnects that guide microwave signals between layers while minimizing radiative loss and suppressing unwanted modes. The electromagnetic environment *between* the chips can be structured using metamaterial principles to precisely control coupling while minimizing crosstalk. Recent work has demonstrated the viability of this approach, developing a 3D integration process based on indium-bump bonding that yields qubits with coherence and gate fidelities approaching state-of-the-art planar devices, paving the way for scalable, multi-chip quantum processors[278,430].

## 9.3 Metamaterials for Phonon Engineering and Suppressing Defect Loss

Beyond their role in controlling electromagnetic fields for connectivity and circuit design, metamaterials present a compelling framework for tailoring the acoustic environment in quantum processors. This approach directly targets one of the most persistent sources of decoherence in superconducting qubits: dielectric loss stemming from microscopic TLSs embedded in amorphous oxides and material interfaces. The prevailing model attributes this loss to a two-stage mechanism, wherein a qubit first excites a TLS, which subsequently relaxes by emitting a phonon into the substrate. Owing to the much higher density of phonon states at microwave frequencies compared to photons, this phonon-mediated pathway dominates TLS relaxation. As a result, even a sparse population of TLSs can severely constrain qubit coherence times.

Phononic metamaterials, structures engineered to manipulate sound wave propagation via periodic patterning[431], can fundamentally disrupt this process. By introducing a phononic bandgap, these metamaterials suppress the density of states (DOS) for phonons within a targeted frequency range. TLSs with transition frequencies lying inside this bandgap lose access to their dominant decay channel, effectively freezing their relaxation and significantly extending their $T_1$ times. This concept has now been experimentally validated with impressive results, showing that acoustic bandgap engineering can shift TLSs from being a decoherence source to a potentially useful quantum degree of freedom. Figure 5 showcases two pivotal studies that exemplify both the strengths and current constraints of this strategy.

In the first study, Chen et al. focused on mitigating decoherence at the most sensitive location of a transmon qubit, its Josephson junctions[183]. They fabricated a flux-tunable transmon on a suspended silicon membrane etched into a two-dimensional phononic crystal (Fig. 5a), placing the junctions within a ~1.2 GHz acoustic bandgap centered near 5.1 GHz. Using the qubit to spectroscopically probe TLSs, they distinguished two populations (Fig. 5b): TLSs with transition frequencies outside the gap ("family A") exhibited typical microsecond-scale lifetimes, while those inside the bandgap ("family B") displayed lifetimes surpassing 5 milliseconds. This dramatic enhancement tracked closely with the simulated suppression of the phononic DOS (Fig. 5c), affirming that spontaneous phonon emission dominates TLS decay and can be inhibited through phononic bandgap design.

Yet the overall coherence of the qubit itself showed only modest improvement. The explanation lies in the remaining weakly-coupled TLSs distributed across other material interfaces, which collectively dominate the loss via the high-density "Fermi limit." To address this, a second study by Odeh et al. embedded the entire transmon, including its capacitor, SQUID loop, and wiring, into a larger phononic crystal structure[432] (Fig. 5d). Rather than focusing on isolated TLSs, they probed the collective behavior of the TLS bath using a hole-burning protocol: the qubit was used to selectively pump TLSs at a given frequency and its thermalization was then monitored. Their results revealed a dramatic shift in dynamics. When tuned inside the bandgap, the qubit exhibited slowed relaxation and a clear biexponential decay profile (Fig. 5e), a hallmark of non-Markovian behavior. This indicated coherent energy exchange between the qubit and a long-lived, structured TLS bath, made possible only because the bandgap extended lifetimes across the ensemble.

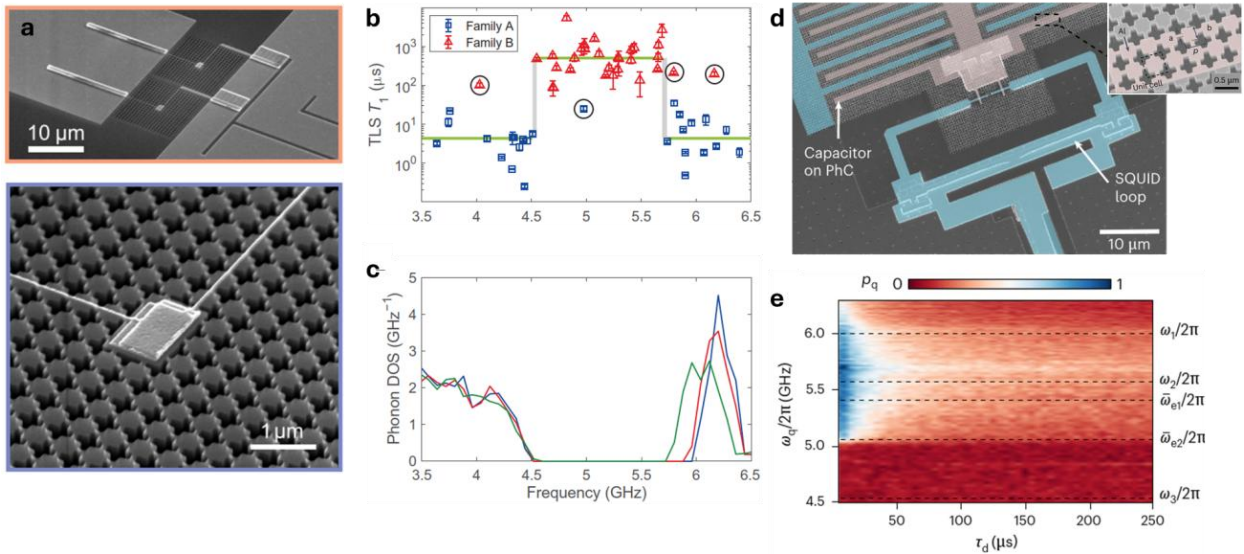

**Figure 5.** Phononic-bandgap engineering stretches TLS lifetimes and drives a superconducting qubit into the non-Markovian regime. (a) Scanning-electron micrograph of a flux-tunable transmon fabricated on a suspended silicon membrane patterned as a two-dimensional phononic crystal; the interdigitated capacitor (orange frame) sits entirely on the crystal, while the junction island (blue inset) is tethered through nine unit cells that open a ~1 GHz acoustic bandgap. (b) Scatter plot of individual two-level-system (TLS) energy-relaxation times $T_1$ versus transition frequency gathered from seven nominally identical devices. TLSs cluster into a short-lived "family A" and a long-lived "family B", the latter appearing only inside the 4.5–5.7 GHz bandgap (green bar), with lifetimes exceeding 0.5 ms. (c) Finite-element simulations of the phonon density of states (DOS) for three representative unit-cell variations reproduce the measured gap and explain the bifurcation in TLS statistics. Adapted from Chen et al., Sci. Adv. 10, eado624 (2024). Copyright 2024 AAAS, Creative Commons[183]. (d) False-coloured optical micrograph of a separate experiment in which the entire qubit – capacitor, SQUID loop and control wiring – is embedded in a larger phononic metamaterial that shares the same bandgap. (e) Colour map of the qubit excited-state population $p_e$ after a hole-burning sequence as a function of qubit frequency $\omega_q/2\pi$ and delay $\tau_d$. Within the bandgap (dashed lines) population relaxes slowly and shows clear biexponential decay, a hallmark of qubit–TLS back-action and hence non-Markovian dynamics. Adapted from Odeh et al., Nat. Phys. 21, 406 (2025). Copyright 2025 Springer Nature, Creative Commons[432].

### 9.4 The Road Ahead: Hybrid Systems and the Search for New Physics

The future of quantum technologies will likely be defined by the development of increasingly sophisticated *hybrid systems* that combine the strengths of different physical platforms[432]. Superconducting metamaterial circuits, with their engineered electromagnetic properties, serve as an ideal backbone for integrating and controlling other quantum degrees of freedom.

- **Hybridizing with 2D Materials:** The integration of 2D van der Waals (vdW) materials – spanning a vast range of phenomena from unconventional superconductivity to magnetism and topology – with superconducting resonators is a particularly exciting frontier. By coupling a vdW flake to a high-Q resonator, the circuit acts as a highly sensitive probe of the material's microwave properties, complementing traditional transport measurements. Conversely, the unique properties of vdW materials, such as the record-high kinetic inductance of clean vdW superconductors, could be harnessed to build more compact and coherent quantum circuits. The primary challenge remains the fabrication of pristine, low-loss interfaces between these disparate material systems.

- **Coupling to Mechanical and Phononic Systems:** Hybrid systems coupling superconducting circuits to nanomechanical oscillators are enabling the field of *quantum acoustodynamics*. Here, phonons (quantized vibrations) can be used as a quantum bus to mediate long-range interactions between qubits or to transduce quantum information between microwave and optical domains[433–438]. Furthermore, phononic metamaterials can be designed with acoustic bandgaps to shield qubits from vibrational noise and phonon-mediated decay channels, offering a new route to enhancing coherence[183,439].

The sheer complexity of the design space for these advanced metamaterial and hybrid systems necessitates new design tools. Here, *artificial intelligence and machine learning (AI/ML)* are emerging as powerful techniques for the inverse design of quantum devices[440–443]. Traditional design involves proposing a geometry, simulating its properties, and iterating. Inverse design flips this process: by specifying a desired electromagnetic response or target Hamiltonian, machine learning algorithms can explore a vast parameter space to discover non-intuitive device geometries that achieve the target functionality. This can be framed as an optimization problem, minimizing a loss function $L(\theta)$ that quantifies the difference between the target Hamiltonian $H_{\text{target}}$ and the simulated Hamiltonian $H_{\text{sim}}(\theta)$ for a device with geometric parameters $\theta$: $\|H_{\text{target}} - H_{\text{sim}}(\theta)\|^2$. This approach has the potential to dramatically accelerate the development of novel quantum devices with optimized performance.

Finally, the field will continue to be a fertile ground for exploring fundamental physics. The ability to engineer Hamiltonians with high precision allows these circuits to act as quantum simulators, probing physical regimes that are inaccessible in naturally occurring systems[444–447]. The exploration of exotic phenomena like synthetic gauge fields in loop-coupled cavity-magnon systems[444], the quantum dynamics of skyrmions[447,448], and the realization of topological phases of matter will not only deepen our understanding of quantum mechanics but will also inspire the next generation of quantum technologies. The Jaynes-Cummings-Hubbard model, realized as an array of coupled cavities and qubits, remains a canonical example, allowing for the direct simulation of quantum phase transitions and many-body physics with light. The continued synthesis of electromagnetic theory, materials science, and advanced fabrication, all guided by the principles of metamaterial design, is paving the way for a future of scalable, fault-tolerant quantum computers built upon a foundation of engineered quantum matter.

## 10. Conclusion

The journey towards a fault-tolerant quantum computer is fundamentally a challenge of mastering the electromagnetic environment where quantum information is stored and processed. Throughout this review, we have established that superconducting metamaterials provide a uniquely powerful and versatile toolkit to achieve this mastery. We began by outlining the foundational synergy between the low-loss, high-inductance properties of superconductors and the design freedom of metamaterial engineering, showing how these principles enable the explicit construction of bespoke quantum Hamiltonians. By arranging simple circuit elements into complex arrays, it is possible to sculpt the photonic dispersion and impedance in ways that directly address the primary obstacles of decoherence, connectivity, and control scalability.

While practical hurdles in materials science and fabrication persist, particularly the mitigation of dielectric loss from two-level systems, the field has achieved remarkable progress, pushing resonator quality factors and qubit coherence times to record levels. The true power of this paradigm, however, lies in moving beyond incremental improvements. By designing environments that support topologically protected states, non-radiating anapole modes, or entirely new qubit paradigms, metamaterials offer pathways to building intrinsic error resilience directly into the hardware.

In essence, metamaterials resolve the tension between processor complexity and quantum performance. They enable the hardware-efficient bosonic codes and all-to-all connectivity crucial for advanced quantum algorithms, transforming the very architecture of quantum processors from a static backplane into an active, computational fabric. The deliberate engineering of these artificial quantum materials represents one of the most promising avenues toward realizing the transformative potential of large-scale quantum computation.

## AUTHOR DECLARATIONS

### Conflict of Interest

The authors have no conflicts to disclose.

### Author Contributions

**Alex Krasnok:** Conceptualization; Formal analysis; Investigation; Methodology; Project administration; Resources; Supervision; Validation; Writing – original draft; Writing – review & editing.

## DATA AVAILABILITY

The data that support the findings of this study are available from the corresponding author upon reasonable request.

[92] E.T. Jaynes, and F.W. Cummings, "Comparison of Quantum and Semiclassical Radiation Theories with Application to the Beam Maser," Proc. IEEE **51**(1), 89–109 (1963).

[93] A. Kavokin, J.J. Baumberg, G. Malpuech, and F.P. Laussy, *Microcavities* (Oxford University Press, 2007).

[94] J. Zhang, and Y. Jiang, "Quantum phase diagrams of the Jaynes–Cummings Hubbard models in non-rectangular lattices," Laser Phys. **27**(3), 035203 (2017).

[95] S. Schmidt, and G. Blatter, "Strong Coupling Theory for the Jaynes-Cummings-Hubbard Model," Phys. Rev. Lett. **103**(8), 086403 (2009).

[96] J.-L. Ma, B. Liu, Q. Li, Z. Guo, L. Tan, and L. Ying, "Many-body phases in Jaynes-Cummings-Hubbard arrays," Phys. Rev. A **109**(3), 033707 (2024).

[97] K. Toyoda, Y. Matsuno, A. Noguchi, S. Haze, and S. Urabe, "Experimental Realization of a Quantum Phase Transition of Polaritonic Excitations," Phys. Rev. Lett. **111**(16), 160501 (2013).

[98] A.D. Greentree, C. Tahan, J.H. Cole, and L.C.L. Hollenberg, "Quantum phase transitions of light," Nat. Phys. **2**(12), 856–861 (2006).

[99] J. Koch, and K. Le Hur, "Superfluid–Mott-insulator transition of light in the Jaynes-Cummings lattice," Phys. Rev. A **80**(2), 023811 (2009).

[100] D.G. Angelakis, M.F. Santos, and S. Bose, "Photon-blockade-induced Mott transitions and X Y spin models in coupled cavity arrays," Phys. Rev. A **76**(3), 031805 (2007).

[101] D. Rossini, and R. Fazio, "Mott-Insulating and Glassy Phases of Polaritons in 1D Arrays of Coupled Cavities," Phys. Rev. Lett. **99**(18), 186401 (2007).

[102] M.J. Hartmann, F.G.S.L. Brandão, and M.B. Plenio, "Strongly interacting polaritons in coupled arrays of cavities," Nat. Phys. **2**(12), 849–855 (2006).

[103] N. Na, S. Utsunomiya, L. Tian, and Y. Yamamoto, "Strongly correlated polaritons in a two-dimensional array of photonic crystal microcavities," Phys. Rev. A **77**(3), 031803 (2008).

[104] J. Figueroa, J. Rogan, J.A. Valdivia, M. Kiwi, G. Romero, and F. Torres, "Nucleation of superfluid-light domains in a quenched dynamics," Sci. Rep. **8**(1), 12766 (2018).

[105] M. Göppl, A. Fragner, M. Baur, R. Bianchetti, S. Filipp, J.M. Fink, P.J. Leek, G. Puebla, L. Steffen, and A. Wallraff, "Coplanar waveguide resonators for circuit quantum electrodynamics," J. Appl. Phys. **104**(11), 113904 (2008).

[106] F. Mallet, F.R. Ong, A. Palacios-Laloy, F. Nguyen, P. Bertet, D. Vion, and D. Esteve, "Single-shot qubit readout in circuit quantum electrodynamics," Nat. Phys. **5**(11), 791–795 (2009).

[107] J. Majer, J.M. Chow, J.M. Gambetta, J. Koch, B.R. Johnson, J.A. Schreier, L. Frunzio, D.I. Schuster, A.A. Houck, A. Wallraff, A. Blais, M.H. Devoret, S.M. Girvin, and R.J. Schoelkopf, "Coupling superconducting qubits via a cavity bus," Nature **449**(7161), 443–447 (2007).